\documentclass[
aip,cha,
 amsmath,amssymb,
 reprint,%
]{revtex4-1}

\usepackage{mathptmx}
\usepackage{etoolbox}

\DeclareMathSizes{12}{17.28}{9}{7}

\usepackage{float}
\usepackage{dcolumn}% Align table columns on decimal point
\usepackage{bm}% bold math
\usepackage{bbold}
\usepackage[multidot]{grffile}
\usepackage[colorlinks=true, breaklinks=true]{hyperref}% add hypertext capabilities
\usepackage[mathlines]{lineno}% Enable numbering of text and display math
\DeclareMathOperator{\dd}{d\!} % pour avoir un d de dérivée droit
\usepackage{xcolor}

\usepackage{xspace}

\usepackage[utf8]{inputenc}
\usepackage{tikz}
\usetikzlibrary{shapes.geometric,fit,arrows,babel}
\usepackage[utf8]{inputenc} % allow utf-8 input
\usepackage[T1]{fontenc}    % use 8-bit T1 fonts
\usepackage{hyperref}
\hypersetup{colorlinks,
citecolor=blue
}
\usepackage{url}            % simple URL typesetting
\usepackage{booktabs}       % professional-quality tables
\usepackage{amsfonts}       % blackboard math symbols
\usepackage{nicefrac}       % compact symbols for 1/2, etc.
\usepackage{microtype}      % microtypography
\usepackage{lipsum}
\usepackage{amsmath,amssymb,amsthm}
\usepackage[normalem]{ulem}
\usepackage[colorinlistoftodos,prependcaption,textsize=tiny]{todonotes}
\usepackage{graphicx}
\usepackage{wrapfig}
\definecolor{Agreen}{rgb}{0.1, 0.6, 0.1} % define new green
\usepackage{multirow}

\usepackage[linesnumbered,ruled,vlined]{algorithm2e}
\usepackage[noend]{algpseudocode}
\SetKwInput{KwInput}{Input}                % Set the Input
\SetKwInput{KwOutput}{Output}              % set the Output

\SetCommentSty{mycommfont}

\newcommand{\marius}[1]{\textcolor{black}{ #1} }
\newcommand{\Jinjie}[1]{\textcolor{black}{ #1} }

\makeatletter
\def\@email#1#2{%
 \endgroup
 \patchcmd{\titleblock@produce}
  {\frontmatter@RRAPformat}
  {\frontmatter@RRAPformat{\produce@RRAP{*#1\href{mailto:#2}{#2}}}\frontmatter@RRAPformat}
  {}{}
}%
\makeatother
\begin{document}

\preprint{AIP/123-QED}

\title[Inverse stochastic resonance in adaptive small-world neural networks]{Inverse stochastic resonance in adaptive  small-world neural networks}
% Force line breaks with \\
\author{Marius E. Yamakou}
\email{marius.yamakou@fau.de}
\affiliation{Department of Data Science, Friedrich-Alexander-Universit\"{a}t Erlangen-N\"{u}rnberg, Cauerstr. 11, 91058 Erlangen, Germany}

\author{Jinjie Zhu}
\email{jinjiezhu@nuaa.edu.cn}
 \affiliation{State Key Laboratory of Mechanics and Control for Aerospace Structures, College of Aerospace Engineering, Nanjing University of Aeronautics and Astronautics, Nanjing 210016, China}
 
\author{Erik A. Martens}
\email{erik.martens@math.lth.se}
\affiliation{Centre for Mathematical Sciences, Lund University, S\"{o}lvegatan 18B, 221 00 Lund, Sweden}
\affiliation{Center for Translational Neurosciences, University of Copenhagen, Blegdamsvej 3, 2200 Copenhagen, Denmark}

\date{\today}% It is always \today, today,
             %  but any date may be explicitly specified

\begin{abstract}
\marius{Inverse stochastic resonance (ISR) is a counterintuitive phenomenon where noise reduces the oscillation frequency of an oscillator to a minimum occurring at an intermediate noise intensity, and sometimes even to the complete absence of oscillations.} \marius{In neuroscience,} ISR was first experimentally verified with cerebellar Purkinje neurons [A. Buchin \textit{et al.}, PLOS Computational Biology 12, e1005000 (2016)]. These experiments showed that ISR enables a locally optimal information transfer between the input and output spike train of neurons. 
Subsequent studies have further demonstrated the efficiency of information processing and transfer in neural networks with small-world network topology.
We have conducted a numerical investigation into the impact of adaptivity on ISR in a small-world network of noisy FitzHugh-Nagumo (FHN) neurons, operating in a \marius{bi-meta-stable regime consisting of a meta-stable fixed point and a meta-stable limit cycle.} Our results show that the degree of ISR is highly dependent on the value of the FHN model's timescale separation parameter $\varepsilon $.
The network structure undergoes dynamic adaptation via mechanisms of either spike-time-dependent plasticity (STDP) with potentiation-/depression-domination parameter $P$, or homeostatic structural plasticity (HSP) with rewiring frequency $F$.
We demonstrate that both STDP and HSP amplify the effect of ISR when $\varepsilon $ lies within the bi-stability region of FHN neurons.
Specifically, at larger values of $\varepsilon $ within the bi-stability regime, higher rewiring frequencies $F$ are observed to enhance ISR at intermediate (weak) synaptic noise intensities,  while values of $P$ consistent with depression-domination (potentiation-domination) consistently enhance (deteriorate) ISR.
Moreover, although STDP and HSP control parameters may jointly enhance ISR, $P$ has a greater impact on improving  ISR compared to $F$. Our findings inform future ISR enhancement strategies in noisy artificial neural circuits, aiming to optimize local information transfer between input and output spike trains in neuromorphic systems, and prompt venues for experiments in neural networks.
\end{abstract}

\maketitle

\begin{quotation}
The impact of noise on nonlinear dynamical systems often yields counter-intuitive behaviors, such as the stabilization of otherwise unstable deterministic states, the \Jinjie{suppression}, or enhancement of oscillation coherence. Classic examples of stochastic \marius{facilitation} include phenomena like stochastic resonance (SR) and coherence resonance (CR). In SR, adding (an optimal intensity of) noise to a nonlinear bi-stable system (or, systems with sensory thresholds, such as neurons) enhances its response to weak external periodic signals, making imperceptible signals detectable \cite{longtin1993stochastic}. Experimental research has shown that SR maximizes information flow in sensory neurons at an optimal noise level \cite{nozaki1999effects}. CR, on the other hand, does not require an external periodic signal \cite{pikovsky1997coherence} and occurs when the regularity of noise-induced oscillations in an excitable system is a peaked (non-monotonic) function of the noise amplitude, with oscillators being optimally correlated at a certain non-zero noise intensity. CR has also been studied experimentally \cite{pisarchik2023coherence}. 
More recent studies of stochastic individual neurons and neural network models have identified another form of non-linear response to noise, inverse stochastic resonance (ISR)~\cite{gutkin2009inhibition,tuckwell2009inhibition}, an effect which later also has been observed experimentally in neurons~\cite{Buchin2016PlosCB} \marius{and electroconvection in liquid crystal \cite{Huh2016inverse,huh2023control}.} ISR leads to an inverse resonance, i.e., the response curve possesses a trough (minimum),  unlike SR and CR which result in a response curve with a peak (maximum); or even the \marius{complete quenching of noise perturbed oscillations.}  \Jinjie{While the effect of network adaptivity on SR and CR has been investigated
extensively, less attention has been given to the case of ISR in adaptive networks \cite{bavcic2018inverse,bavcic2020two}. To address this gap,} we study ISR in adaptive small-world neural networks, with two key mechanisms of adaptivity: STDP and HSP. Our study shows that these adaptive mechanisms strengthen ISR, thus offering insights into the role and control of ISR in neural and other complex systems, and offering guidance for future experimental investigations.
\end{quotation}

\section{\label{Sec. I}Introduction}

Noise-induced resonances are counter-intuitive phenomena where the introduction of stochastic fluctuations, or noise, into a nonlinear dynamical system modulates the system's activity and response. Stochastic resonance (SR) is a well-documented phenomenon where noise enhances a nonlinear system's response to weak periodic signals and has been extensively studied and leveraged across various fields particularly in neuroscience~\cite{hanggi2002stochastic,longtin1993stochastic,vazquez2017stochastic,gluckman1998stochastic,bulsara1991stochastic}. For coherence resonance, the addition of a specific amount of noise in excitable system renders oscillatory responses most coherent~\cite{pikovsky1997coherence}, and thereby optimizes signal detection and processing.  Inverse stochastic resonance (ISR) is a counterpart phenomenon, since stochastic fluctuations or noise results in the reduction or even \marius{quenching of oscillations.} Thus, ISR conceptually represents a paradigm shift in our understanding of noise-induced behaviors in complex systems, as it displays that noise \Jinjie{does not only play} an excitatory but also an inhibitory role~\cite{tuckwell2009inhibition}. 

ISR was first identified in the context of neuronal dynamics with the observation that certain neurons, subjected to an optimal level of noise, could decrease and sometimes even completely quench their mean firing rate into quiescence~\cite{gutkin2009inhibition}. The experimental validation of ISR in both biological~\cite{Buchin2016PlosCB} and physical~\cite{huh2023control} systems underscores its relevance in the real world. By elucidating how noise can modulate neural activity, ISR provides insights into the delicate balance between excitation and inhibition in the brain. The discovery of ISR has significant implications for understanding both the function and dysfunction of the brain, especially in terms of regulatory mechanisms of neural circuits. For example, it is generally believed that neurons convey information via spiking interactions. Consequently, the occurrence of ISR can on one hand be viewed as a constraint on information processing in neural systems. On the other hand, ISR could also be crucial for computational processes that require diminished firing activity without chemical inhibitory neuro-modulation, or for those processes requiring intermittent bursts of activity \cite{paydarfar2006noisy}. \marius{Furthermore, ISR may be beneficial in neuronal systems, including suppression of pathologically long short-term memories \cite{Uzuntarla2013dynamical, 2017double,Buchin2016PlosCB,Schmerl2013Channel} and the emergence of up-down states \cite{vyazovskiy2013sleep}.}  Thus, the presence of ISR could be beneficial in many scenarios. 
Moreover, the quenching effect of ISR holds therapeutic potential, where controlled noise might be used to reduce excessive neural activity and prevent conditions marked by hyper-excitability, such as epilepsy~\cite{buchin2015modeling}. Beyond neuroscience, the intriguing ISR effect is also relevant to other systems, \marius{such as electroconvection in liquid crystal}~\cite{huh2023control, Huh2016inverse} and ecological systems~\cite{touboul2018complex}.

\marius{As we mentioned earlier, from a dynamical systems perspective, ISR can emerge through various mechanisms, including bi-stability in the vicinity of a sub-critical Hopf bifurcation, \Jinjie{in which case a biased switching between the oscillatory and the quasi-stationary meta-stable states derived from the deterministic attractors of system induces ISR \cite{gutkin2009inhibition,tuckwell2009inhibition}.}  The switching rates between the two meta-stable \Jinjie{states} become notably asymmetric at an intermediate noise level, with the system spending substantially more time in a quasi-stationary fixed point~\cite{Torres2020theoretical,uzuntarla2017inverse}. This results in a distinctive non-monotonic relationship between the frequency of oscillations and noise, a hallmark of ISR.  In the current paper, we focus on ISR induced via this type of bi-stability.
 We emphasize that such a bi-stability is not the only mechanism that can lead to ISR. It has been shown that there are other scenarios, related to, for example, phase-sensitive excitability of a limit cycle \cite{Igor2018phase} and the noise-enhanced stability of an unstable fixed point \cite{bavcic2020two,zhu2021unified}, which do not require bi-stability.}
 
Recent years have shown a surge of interest in exploring various aspects of the ISR phenomenon. One investigation~\cite{guo2011inhibition} delved into the influence of temporal noise correlations on ISR, revealing that colored noise exerts a more significant suppressive impact on neural activity when compared to Gaussian white noise. 
Another study~\cite{tuckwell2011effects} scrutinized ISR in a more realistic setting including spatial extension, demonstrating that \marius{intermediate}  noise can impede spiking when signal and noise inputs coincide spatially on the neuron; vice versa, if the signal and noise are unevenly distributed, the noise does not disrupt spiking activity, irrespective of the neuron's extension on a spatial domain. 
A further study~\cite{2017double} found that ISR can emerge in static networks as a consequence of a variety of factors, including channel noise, connection strength, synaptic currents with excitatory and inhibitory terms, and topological features of the network, e.g., degree distribution and mean connectivity degree.

A significant number of studies investigated ISR in individual neurons and neural networks as a function of various types of noise~\cite{wang2022non,lu2020inverse,zhao2019levy}, spatial extension of the neuron model~\cite{tuckwell2011effects,liu2024effects,zhang2021autapse}, electromagnetic induction due to ions moving in the neuronal  membrane~\cite{ye2023inverse}, conductance-driven input~\cite{tuckwell2009inhibition}, neuronal morphology~\cite{liu2024effects}, time delays and coupling strength~\cite{liu2024effects,zhang2021autapse}, electrical synapses versus (inhibitory and excitatory) chemical synapses~\cite{2017double}, electrical and chemical autapses (i.e., time-delayed synaptic connections where a neuron forms a synapse with itself)~\cite{zhang2021autapse}, the average degree of scale-free neural network size~\cite{2017double}. 
\marius{Despite the large range of scenarios explored in these research efforts, the question of how adaptivity in oscillator networks affects ISR has received insufficient attention \cite{bavcic2018inverse,bavcic2020two,2017double}.} 

\marius{In Ref. \cite{bavcic2018inverse}, it has been shown that adaptive coupling mimicking Hebbian-like plasticity in excitable active rotators has facilitatory effects on the occurrence of ISR.}
\Jinjie{In  Ref. \cite{bavcic2020two}, ISR is studied in two adaptively coupled stochastic active rotators and showed two distinct mechanisms for ISR: noise-enhanced stability of an unstable fixed point and biased switching between meta-stable states.
}
The study \Jinjie{in Ref. \cite{Uzuntarla2013dynamical}} examined ISR by analyzing how a single neuron with synaptic dynamics is affected by the independent (uncorrelated) spiking activity of numerous other neurons. In this setup, presynaptic neurons are modeled as independent Poisson spike generators emitting uncorrelated spikes with a certain frequency. For synaptic transmission to the postsynaptic neuron, the authors adopt the dynamic synapse formulation of Tsodyks and Markram~\cite{tsodyks1998neural}. 
It was shown that dynamic synapses featuring short-term synaptic plasticity may expand or shrink the interval of presynaptic firing rate over which ISR in the single postsynaptic neuron is present.
Examining short-term depression and facilitation of the noisy postsynaptic current, other authors~\cite{2017double} found that double inverse stochastic resonance (DISR) can occur with two distinct dips at different presynaptic firing rates. However, this research 
fails to incorporate a large range of other crucial plasticity principles that regulate the adaptability of neural networks in the brain. \marius{Thus, the current study aims to narrow the gap in the literature regarding how ISR in adaptive neural networks is influenced by both spike-timing-dependent plasticity (STDP) and homeostatic structural plasticity (HSP).}

\marius{It is worth pointing out that in neurobiology, plasticity can be classified into two types: short-term (transient) and long-term plasticity. Short-term synaptic plasticity (STP)~\cite{tsodyks1998neural,2017double}  is an example of transient plasticity, whereas HSP~\cite{bennett2018rewiring,van2017rewiring,yamakou2023combined} and STDP~\cite{gerstner1996neuronal,markram1997regulation} are examples of long-term plasticity.}
In particular, STP and STDP are two distinct mechanisms that modulate synaptic strength, but they operate on different timescales, under different principles, and have different functional roles. In terms of timescale, STDP involves long-lasting changes, while STP involves transient changes. In terms of dependence on timing, STDP is highly dependent on the exact timing of pre-and postsynaptic spikes, while STP depends more on the recent history of synaptic activity and involves mechanisms such as neurotransmitter release probability. In terms of function, STDP is primarily associated with learning and memory, encoding long-term changes, whereas STP is involved in modulating synaptic transmission over short periods, affecting real-time signal processing. Synaptic plasticity in neural networks denotes the ability to adjust the potency of synaptic links over time and/or transform the neural network's structural configuration according to certain principles. Two primary mechanisms linked to adaptive regulations in neural networks are spike-time-dependent plasticity (STDP) and homeostatic plasticity (HSP). Synaptic modifications induced by STDP hinge on the repeated pairing of pre-and postsynaptic membrane potentials, with the extent and direction of these changes contingent on the precise timing of the neuronal firing. The exact timing of pre-and postsynaptic spikes determines whether synaptic weights undergo long-term depression (LTD) or long-term potentiation (LTP), which correspond to a lasting decrease or increase in synaptic strength, respectively~\cite{gerstner1996neuronal,markram1997regulation}.

Synaptic modifications induced by HSP entail altering neuronal connectivity by creating, pruning, or rearranging synaptic connections. \marius{This leads to modifications in the network's architecture while preserving its operational framework, thereby enhancing the specialized functions of interconnected neuronal groups and enhancing the efficiency of sensory processing \cite{shine2016dynamics}. Initial indications of structural plasticity were identified through histological analyses of spine density in response to new sensory experiences or training~\cite{greenough1988anatomy}. Additional studies revealed that the micro-connectome, which describes the connectome at the level of individual synapses, undergoes rewiring~\cite{bennett2018rewiring,van2017rewiring,yamakou2023combined}. Although brain networks conform to distinct topologies like small-world and random networks~\cite{hilgetag2016brain, valencia2008dynamic} and exhibit dynamic behavior over time, recent research indicates that these networks can enhance information processing efficiency through homeostasis~\cite{butz2014homeostatic}. 
Motivated by these studies,  this paper examines ISR in a 
time-varying small-world network of FitzHugh-Nagumo neurons evolving via STDP while adhering to its small-worldness via HSP at all times.}

It should be noted that in contrast to the well-known phenomenon of homeostatic plasticity, where the 
neurons' spiking rate is kept relatively constant with changing excitations \cite{turrigiano2004homeostatic}, the phenomenon of homeostatic \emph{structural} plasticity (HSP) used in this work involves the mechanism via which \Jinjie{the statistical features of the network are preserved} despite the rewiring of synapses. Thus, our HSP rule is independent of spike activity in the current study.

The main objectives of this study are the following.  
First, we examine how ISR is influenced 
by $\varepsilon $, i.e., the timescale separation between the fast membrane potential and the slow recovery current variables of the neuron model; specifically, we investigate how the \marius{bi-stability, which is in our scenario necessary for the emergence of ISR,} is affected by $\varepsilon $. Second, we study how STDP and \marius{HSP} change the non-monotonic mean-firing rate response that is characteristic of ISR while varying the STDP control parameter $P$ (which determines whether STDP induces potentiation or depression-domination average synaptic weight) and the HSP control parameter $F$ (which determine how quickly the synapses of the small-world network architecture \marius{rewire} while maintaining its small-world properties, \marius{i.e., high clustering
coefficient and a short characteristic path length)}. 
To investigate these issues, we employed systematic and extensive
numerical simulations.
\section{\label{Sec. II}Model}
\subsection{Neuron model}
We consider a paradigmatic model with well-known biological relevance, the FitzHugh-Nagumo (FHN) neuron model~\cite{fitzhugh1960thresholds,fitzhugh1961impulses,nagumo1962active}:
\begin{eqnarray}\label{eq:1}
\left\{\begin{array}{lcl}
\displaystyle{\frac{dV_i}{dt}}&=&V_i\big(a-V_i\big)\big(V_i-1\big)-W_i+I_{i}^\text{syn}(t) + \eta_i(t),\\[3.0mm]
\displaystyle{\frac{dW_i}{dt}}&=&\varepsilon\big(bV_i-cW_i\big).
\end{array}\right.
\end{eqnarray}
where $V_i=V_i(t)\in\mathbb{R}$ and $ W_i=W_i(t)\in\mathbb{R}$ represent the fast membrane potential and slow recovery current variables of the neuron $i=1,\ldots, N$, respectively; $0<\varepsilon \ll1$ defines the timescale separation between $V_i$ and $W_i$;  $a$, $b>0$ and $c > 0$ are parameters changing the dynamic behavior of the neuron.  
The terms $\eta_i$ ($i=1,...,N$) are independent Gaussian noises with zero mean, standard deviation $\sigma$ (noise intensity), and correlation function $\langle \eta_i(t),\eta_j(t') \rangle= \sigma^2\delta_{ij}(t-t')$.

Note that, when we study a \emph{single neuron} ($N=1$), we drop the subscripts $i$ and let the synaptic input $I^\text{syn}=0$.

\subsubsection{Numerical Integration of SDE.}
To integrate the stochastic differential equations (SDEs) Eqs.~\eqref{eq:1} in time, we used the Euler–Maruyama algorithm \cite{higham2001algorithmic} with a small time step $\dd t = 0.0025$ and an integration time of $T=7.0\times10^{3}$ units, which is sufficiently long to overcome transient behavior that \Jinjie{lasts} for $T_0=1.0\times10^{3}$ units.

\subsection{Network model}
To include synaptic interactions in a neural network, we introduce the synaptic input current $I_{i}^\text{syn}(t)$ in Eq.~\eqref{eq:1}, 
which models excitatory uni-directional chemical synapses between the neurons along their synaptic connections. The synaptic input current $I_{i}^\text{syn}(t)$ for the $i$th neuron at time $t$ is given by
\begin{equation}\label{eq:3}
I_{i}^\text{syn}(t) = - \frac{1}{k_i(t)}\sum_{i \neq j=1}^{N}\ell_{ij}g_{ij}s_j(t)\big[V_i(t)- V_\text{syn}\big].
\end{equation}
This term sums the synaptic input currents from all pre-synaptic neurons adjacent to neuron $i$. Such an interaction occurs if the neuron $j$ is pre-synaptic to the neuron $i$, i.e., if the connectivity matrix $L (=\{\ell_{ij}\})$ is $\ell_{ij}=1$ ; otherwise, $\ell_{ij}=0$. Specifically, we consider a small-world (SW) network \cite{bassett2006small,bassett2006adaptive,liao2017small,muldoon2016small} constructed using a Watts-Strogatz network algorithm \cite{watts1998collective,strogatz2001exploring}, where the network's Laplacian matrix $L (=\{\ell_{ij}\})$ 
is a zero-row-sum matrix. The sum over all synaptic inputs is normalized by 
the in-degree of the $i$th neuron (i.e., the number of synaptic inputs to the neuron $i$), $k_i(t)=\sum_{j\neq i}\ell_{ij}=k_i$.
The matrix $g_{ij}$ represents the weight of the connection from the pre-synaptic neuron $j$ to the post-synaptic neuron $i$. Note that the connectivity matrix and the synaptic weights will adapt over time when we introduce plasticity mechanisms (see Sec.~\ref{sec:adaptive_network}).

An input current is 
modulated by the fraction of open synaptic ion channels, $s_j$, in a pre-synaptic neuron $j$. Finally, the membrane potential $V_i$ of the incident neuron $i$ is compared to the \marius{reversal} potential $V_\text{syn}$.

The fraction of open synaptic ion channels at time $t$ of the $j$th neuron is represented by $s_j(t)$ in Eq.~\eqref{eq:3} and evolves in time according to \cite{yu2015spike}:
 \begin{equation}\label{eq:4}
\frac{ds_j}{dt} = \frac{2(1 - s_j)}{1 + \displaystyle{\exp\Bigg[- \frac{V_j(t)}{V_\text{shp}}\Bigg]}}-s_j,
 \end{equation}
 where $V_j(t)$ is the action potential of the pre-synaptic neuron $j$ at time $t$; $V_\text{shp}=0.05$ determines the threshold of the membrane potential above which the post-synaptic neuron $i$ is affected by the pre-synaptic neuron $j$.
 
Our study focuses on the inhibition of spiking activity triggered solely by noise through the effect of ISR. Nevertheless, it is well known that inhibitory synapses can also inhibit spiking activity in neural networks independently of ISR, operating through distinct mechanisms that can manifest both with and without the bi-stability required for ISR. Thus, to avoid the inhibition of spiking activity that inhibitory synapses might induce and focus only on the inhibition of spiking activity induced by the ISR effect, we shall fix the reversal potential $V_\text{syn}$ in Eq.~\eqref{eq:3} at $V_\text{syn}=2.0$ so that the network in Eq.~\eqref{eq:1} is entirely excitatory. Of course, one could study ISR in neural networks with inhibitory synapses, but for the reason given above, we are not interested in that case.

\subsection{Adaptive network model}\label{sec:adaptive_network}
We intend to investigate the behavior of ISR in an adaptive network of $N$ FHN neurons exhibiting two forms of plasticity --- spike-time-dependent plasticity (STDP) and homeostatic structural plasticity (HSP).
Thus,  the connectivity and synaptic weights are functions of time, i.e., $l_{ij}=l_{ij}(t)$ and $g_{ij}=g_{ij}(t)$, which are updated according to the rules of STDP and HSP explained in the following.

\subsubsection{Spike-time-dependent plasticity (STDP)}
The synaptic strength $g_{ij}(t)$ for each synapse is updated according to a nearest-spike pair-based STDP mechanism~\cite{morrison2007spike}. The update rule according to~\cite{xie2018spike} is then implemented as follows:
\begin{eqnarray}\label{eq:5}
%\begin{split}
\left\{\begin{array}{lcl}
 g_{ij}(t + \Delta t) = g_{ij}(t) + \Delta g_{ij},\\[3.0mm]
\Delta g_{ij}=g_{ij}(t)M(\Delta t),
  \\[3.0mm]
M(\Delta t)=
  \left\{
\begin{array}{ll}
\displaystyle{A\exp{(-\lvert\Delta t\rvert/\tau_{a})}\:\:\text{if}~\Delta t>0, }\\[1.0mm]
\displaystyle{- B\exp{(-\lvert\Delta t\rvert/\tau_{b})}\:\:\text{if}~\Delta t<0,}\\[1.0mm]
0 \:\:\text{if}~\Delta t=0.
\end{array} 
\right.
\end{array}\right.
%\end{split}
\end{eqnarray}
This rule updates the synaptic coupling strength $g_{ij}(t)$ multiplicatively via the synaptic modification function $M$, where $\Delta t=t_i -t_j$, $t_i$ and  $t_j$ represent the spiking time of neurons $i$ and $j$.
The amount of synaptic modification $M$ is controlled \marius{by the parameters adjusting the potentiation and depression rates $A$ and $B$, respectively.}  The potentiation and depression temporal windows of the synaptic modification are controlled by $\tau_a$ and $\tau_b$, respectively.

Studies conducted experimentally have shown that the timeframe during which synaptic \marius{depression} occurs aligns closely with that of synaptic \marius{potentiation} \cite{bi1998synaptic,feldman2005map,song2000competitive}. For our model, synaptic potentiation reliably occurs when the post-synaptic spike occurs within a 2.0 time unit window following the pre-synaptic spike, while depression is induced conversely. Thus, the temporal window parameters for potentiation and depression are fixed at the same value, i.e., $\tau_a = \tau_b$ = 2.0.
The same studies have also shown that the ratio of the adjusting depression and potentiation rate parameters determines whether STDP exhibits long-term potentiation (LTP) or long-term depression (LTD). It was shown that STDP is depression-dominated if $P:=B/A > 1$ and if $P:=B/A < 1$, it is potentiation-dominated \cite{bi1998synaptic,feldman2005map,song2000competitive}. 

In the present study, we wish to investigate ISR when STDP is both depression- and potentiation-dominated. We, therefore, keep the depression rate parameter fixed at $B=0.5$ so that the potentiation rate parameter is always given by $A=0.5/P$. 
Thus, we may consider $P$ as the single control parameter so that STDP is depression-dominated when  $P > 1$ \cite{yamakou2023combined,yamakou2023synchronization}, and potentiation-dominated when $P < 1$. 
We vary $P$ in the interval given by $[5.0\times10^{-6},5.0]$.

Furthermore, we wish to prevent (i) unbounded growth; (ii) negative coupling strengths, \marius{as they may give rise to} inhibitory synapses, which we wish to avoid (for the reason given earlier); and (iii) the complete elimination of synapses (i.e., $g_{ij}=0$). To achieve this, we require that $g_{ij}$ remains bounded, i.e., $g_{ij}\in[g_\text{min},g_\text{max}]=[0.5\times10^{-3},0.1\times10^{-2}]$. Here, the maximum synaptic weight $g_\text{max}$  represents the value above which the bi-stability between the stable fixed point and limit cycle disappears and leaves only an unstable fixed point and a stable limit cycle. We choose a small but non-zero $g_\text{min}$ to prevent the complete deletion of synapses while allowing room for synaptic modifications that do not exceed $g_\text{max}$. Moreover, the initial weight of all excitable synapses is normally distributed in the interval $[g_\text{min},g_\text{max}]$, with mean $g_0=0.75\times10^{-3}$ and standard deviation $\sigma_0=0.15\times10^{-3}$.

\subsubsection{Homeostatic structural plasticity (HSP)}

To mimic HSP in the neural network dynamics given by Eqs.~\eqref{eq:1} and ~\eqref{eq:3}, we generate a time-varying small-world network with rewiring probability $\beta\in(0,1)$ that adheres to its small-worldness at all times
during the integration interval. To achieve this, we implement the following process during the rewiring of synapses \cite{yamakou2023combined}:
To build an initial small-world network,
we used the Watts-Strogatz algorithm \cite{watts1998collective,strogatz2001exploring} with rewiring probability of $\beta=0.25$ and average degree of $\langle k \rangle=4$. A synapse between two distant neurons is rewired to one of the neuron's nearest neighbors with probability $(1 - \beta)F\dd t$. If the synapse is already between two nearest neighbors, it is replaced by a synapse to a randomly chosen distant neuron with probability $\beta F \dd t$.  We consider a node $i$ to be distant to node $j$ if $\lvert i-j \rvert >\langle k \rangle$, where $\langle k \rangle$ is the average degree of the original ring network used in the Watts-Strogatz algorithm \cite{watts1998collective,strogatz2001exploring} to generate the initial small-world topology. 

With a small integration time step of $\dd t=0.0025$ and a rewiring probability of $\beta=0.25$, the parameter $F$ determines whether or not the probabilities given by $(1 - \beta)F\dd t$  and  $\beta F \dd t$ are large enough for the original and subsequent small-world networks to rewire as time advances in steps of $\dd t$.
Thus, the parameter $F$ becomes a proxy for the frequency at which the neurons in a small-world network swap their synapses while preserving the network's small-worldness.  In this paper, we call $F$ the characteristic frequency (which we will measure in per second (Hz) to make the probabilities dimensionless) of the time-varying network topology. 

If $F=0$, then $(1 - \beta)F \dd t=0$  and $\beta F \dd t=0$, and none of the synapses will be rewired. Consequently, the small-world network is time-invariant (static network). As soon as $F>0$, $(1 - \beta)F \dd t>0$  and $\beta F \dd t>0$, and there is a non-zero probability that the network rewires and becomes a time-varying network. However, if $F$ is small, the topology will only slowly change over time. As $F$ increases, the network rewires faster because the probabilities $(1 - \beta)F \dd t$  and $\beta F \dd t$ also increase. 
For example, if the probability that a synapse between two distant neurons is rewired to a nearest neighbor of one of the neurons is unity, i.e., $(1-\beta)F \dd t = 1$, then we compute the maximum rewiring frequency of the network as 
$F=1/((1-\beta) \dd t)$, which is $\approx 533$ Hz~\footnote{\marius{
We can use large rewiring frequencies $F$, even exceeding the spiking frequency, because our study employs an HSP rule independent of spike activity and relies solely on network topology. These large 
$F$ values are numerically necessary to ensure the network has a sufficiently high probability of rewiring its synapses. Furthermore, note that large rewiring frequencies such as $F=533$ Hz do not mean that the real brain rewires at such a high frequency. $F$ is just a proxy of the actual rewiring frequency of synapses in a real brain --- higher (lower) $F$ indicate higher (lower) rewiring frequencies in the brain, aiding in qualitative understanding.}} for the value of $\beta=0.25$ and $\dd t=0.0025$ used in our computations. At the same time, the probability that a synapse replaces the synapse between two nearest neighbors to a randomly chosen distant neuron is computed as $\beta F \dd t=0.25\times533\times0.0025\approx0.33$.

Finally, note that we lose the time dependence in the average degree connectivity (average number of synaptic inputs per neuron) $\langle k \rangle$ because neurons would be able to change their neighbors via the rewiring rules, but we require that they do not change the \textit{number} of neighbors. That is, the number of neighbors is always fixed, but the individual neighbors may be swapped over time with a certain probability.

\section{\label{Sec. III}Numerical observations/measurements}

\paragraph{Ensemble averages.} To quantify the dynamic behavior of Eq.~\eqref{eq:1} and to ensure robust statistical results of the neural network, for a fixed parameter, we calculate value ensemble averages over $R=300$  (independent) realizations with random initial conditions and initial small world network structure. 
For each realization of the neural network, initial conditions $(V_i(0),W_i(0))$ of the $i$th neuron ($i = 1,...,N$) were drawn randomly from a uniform distribution within the range covering the basins of attraction of the stable fixed point and the limit cycle, i.e., $V_i(0)\in(-0.5,1)$, $W_i(0)\in(-0.05,0.2)$. 

\paragraph{Measurements.}
Measurements are always made by excluding a transient time of $T_0=1.0\times10^{3}$ units. For each realization, we then calculated the number of spikes $n_{\text{spike},\ell}$ that occur during the remaining $T-T_0$ time units. Spikes were recorded when the membrane potential variable $V_i(t)$ crosses the threshold $V_{\mathrm{th}}=0.25$ from below. 

\marius{An average (over the number of realizations $R$ and time interval $T-T_0$) firing rate, denoted by $\overline{r_i}$ or $\overline{r}$ for each individual neuron $i$ in the network or a single neuron in Eq.~\eqref{eq:1}, respectively,} was calculated as follows:
\begin{equation}\label{eq:3m1}
    \overline{r_i}=\frac{1}{R(T-T_0)}\sum\limits_{\ell=1}^R n_{\text{spike},\ell}.
\end{equation}
The collective mean firing rate $\langle r \rangle$ of the neural network of $N=70$ neurons was then calculated as
\begin{equation}\label{eq:3m2}
\langle r \rangle =\frac{1}{N}\sum\limits_{i=1}^N \overline{r_i}.
\end{equation}

\paragraph{Spike-time-dependent plasticity (STDP).}
 To investigate how the average coupling strength $\langle g_{ij}\rangle$ of the network changes with the STDP parameter, which affects ISR, we averaged the synaptic weights over the entire population and time:
\begin{equation}\label{eq:8}
%\begin{split}
%\left\{\begin{array}{lcl}
\langle g_{ij}\rangle = \displaystyle{\Bigg \langle\frac{1}{N^2}\sum\limits_{i=1}^{N}\sum\limits_{j=1}^{N}g_{ij}(t)\Bigg \rangle_t},
%\end{array}\right.
%\end{split}
\end{equation}
where $\big\langle\cdot\big\rangle_t$ represents the average over the time interval $[T_0,T]$.

In the following section on Results, we use the mean firing rate $\langle r \rangle$ of Eq.~\eqref{eq:3m2} to study the effect of (i) the bi-stability parameter $\varepsilon $, (ii) the noise intensity $\sigma$, (iii) the STDP rule (controlled by the parameter $P$ defined earlier), and (iv) the HSP rule (controlled by the characteristic frequency parameter $F$) on the occurrence ISR. For an example of the control flow used in the simulations, see Appendix in Sec. \ref{Sec. VI}. This flow of control is easily adapted to produce the rest of the simulations presented in this paper.

\section{\label{Sec. IV}Results and discussion}

\subsection{Bifurcation analysis for a single noiseless neuron}

We provide a brief bifurcation analysis for Eq.~\eqref{eq:1} without noise ($\sigma=0$). 
An important goal of this analysis is to pinpoint the conditions on the parameters that enable the co-existence of a stable fixed point and a stable limit cycle, resulting in a bi-stable regime \marius{in our model.}

For the FHN neuron in Eq.~\eqref{eq:1}, there is a unique fixed point $(V_0,W_0)=(0,0)$  if and only if $(a-1)^2/4<b/c$. This fixed point is stable when $-a/\varepsilon <c$ and $a>-b/c$, that is, when $a<0$ is sufficiently close to zero ($-1\ll a<0$), and in the limit $\varepsilon  \to 0$ only for $a\ge 0$.

The $V$-nullcline given by the graph of $W = -V^3 +(a+1)V^2 - aV$ loses normal hyperbolicity at its maximum $V_+$ and minimum $V_-$ points (i.e., the fold points), each located at $V_\pm=(a+1)/3 \pm \sqrt{(a+1)^2/9 - a/3}$. 
Here, it is worth noting that $V_- < V_0=0$ if and only if $a < 0$.

When $c^2<b/\varepsilon $ and $3\varepsilon  c \le a^2-a +1$, we observe a Hopf bifurcation at 
$V_\text{HB}=(a+1)/3-\sqrt{(a+1)^2/9-(a+\varepsilon  c)/3}$. Since $\varepsilon 
c>0$, it is easy to see that $V_-<V_\text{HB}$, and consequently, the Hopf bifurcation $V_\text{HB}$ is to the right
of the minimum $V_-$ of cubic $V$-nullcline, hence on its ascending branch.  

Whenever a fixed point $(V_0,W_0)=(0,0)$ is on the left descending branch of the $V-$nullcline, that is, to the left of the minimum $V_-$, it is stable, and this stability persists a little into the ascending branch for specific choices of the parameter values $a$, $b$, $c$, and $\varepsilon $. For $a<0$, $V_-<V_0=0$, a Hopf bifurcation occurs when  $\varepsilon =-a/c=:\varepsilon _\text{HB}$. Thus, the fixed point $(V_0,W_0)=(0,0)$ loses stability via Hopf bifurcation as $\varepsilon $ decreases. The stable fixed point $(V_0,W_0)$ and an unforced stable limit cycle $\big[\overline{v}(t),\overline{w}(t)\big]$ co-exist as long as $-a/c<\varepsilon $. In other words, we have bi-stability between the fixed point $(V_0,W_0)$  and a limit cycle if and only if $V_- < V_0=0 < -a/c=\varepsilon _\text{HB}$. 

\begin{figure}[htp!]
\centering
\includegraphics[width=\columnwidth]{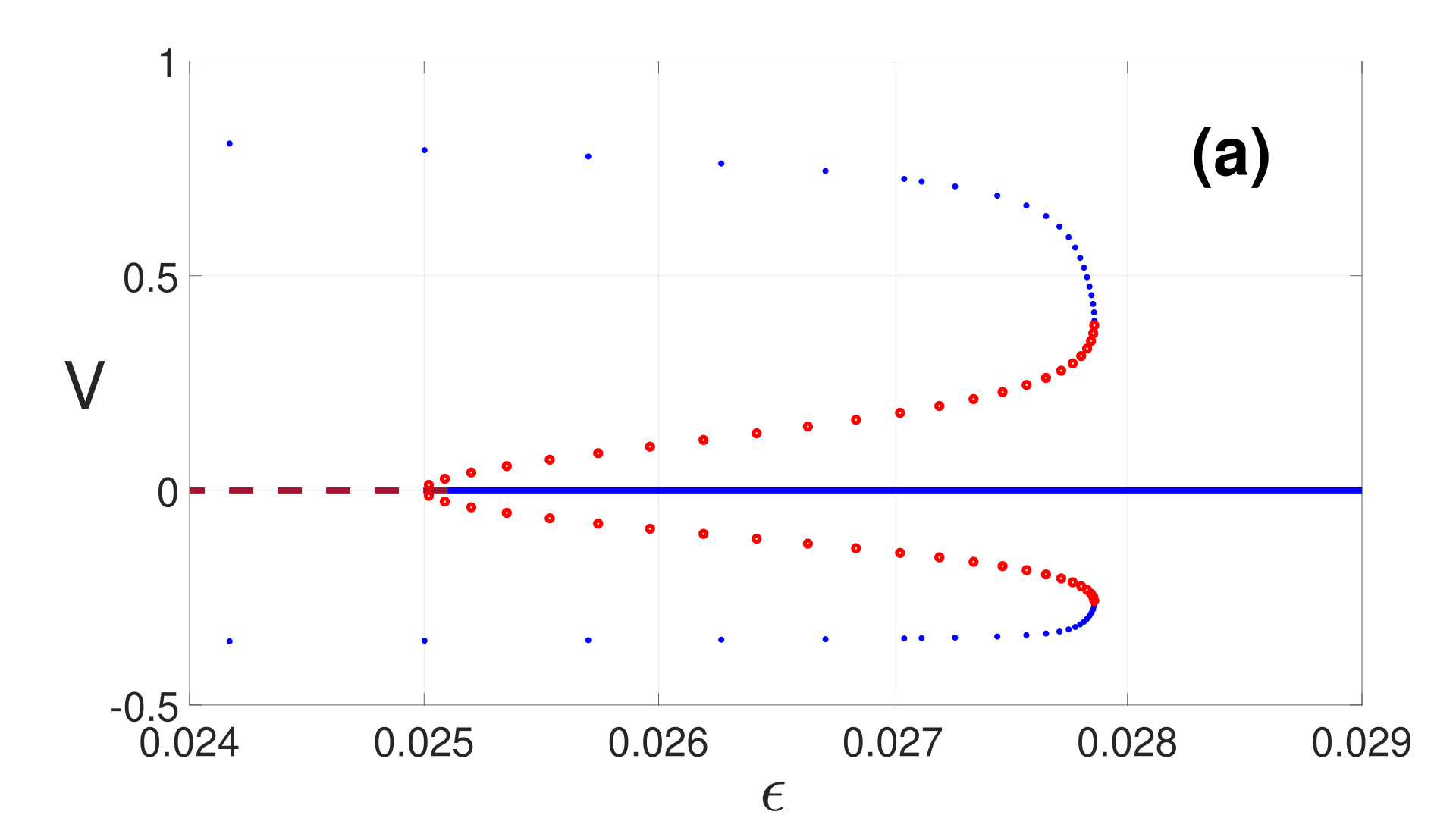}
\includegraphics[width=\columnwidth]{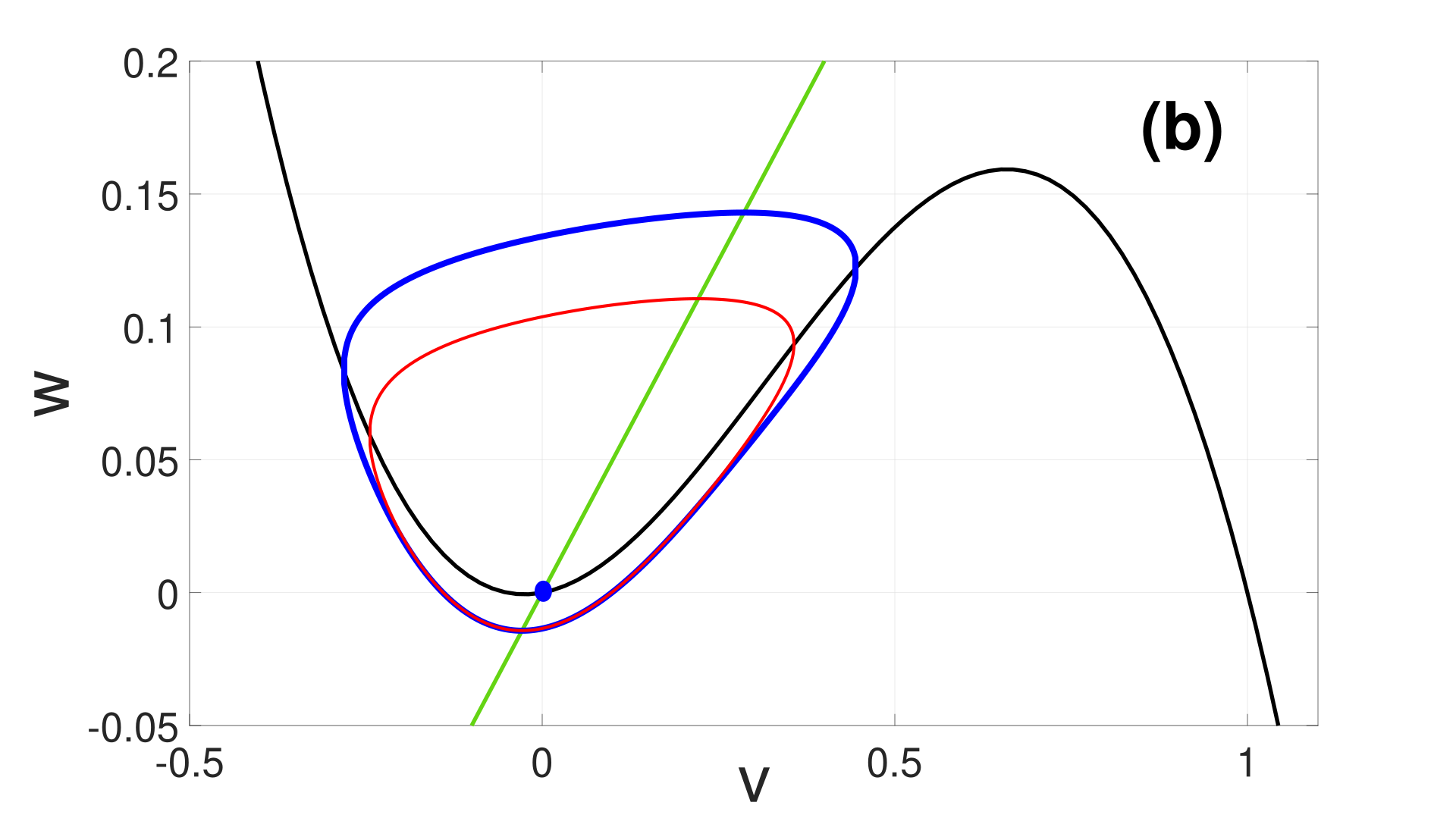}
\caption{\textbf{(a)} Bifurcation diagram for $N=1$ neuron with $I_\text{syn}=0$, Eq.~\eqref{eq:1}.  Membrane potential $V$ vs. timescale parameter $\varepsilon $ of a single noiseless FHN neuron. The bi-stable regime with a \textit{stable} fixed point at $V=0$ (blue line) and \textit{stable} limit cycle (blue dots) resides in the interval $\varepsilon \in[0.025,0.027865)$. An unstable limit cycle (red dots) separates the two stable attractors.
\textbf{(b)} Phase portrait of a single FHN neuron with $\varepsilon =0.02785$ displays the bi-stability between the fixed point at the origin (the unique intersection of the cubic $V$- and $W$-nullclines) and limit cycle (blue). An unstable limit cycle (red) separates the two attractive states. Other parameters:
$a=-0.05$, $b=1.0$, $c= 2.0$, $\sigma=0.0$.}
\label{fig:1}
\end{figure}

Fig.~\ref{fig:1} illustrates the dynamic behavior of a single deterministic FHN neuron, where we chose parameters (see caption) such that bi-stable behavior is present, i.e., $V_-=-0.025244<V_0=0<-a/c=\varepsilon _\text{HB}=0.025>0$.
The bifurcation diagram in panel \textbf{(a)} displays a stable fixed point at $(V_0,W_0)=(0,0)$ and stable limit cycle for the parameter range $\varepsilon \in[\varepsilon _\text{HB},\varepsilon _\text{FB})= [0.025,0.027865)$, where $\varepsilon _\text{FB}$ represents the fold bifurcation point at which the stable and unstable limit cycles coalescence and annihilate, leaving behind only the stable fixed point $(V_0,W_0)=(0,0)$.
The phase portrait in panel \textbf{(b)} shows the $V$- and $W$-nullclines in black and green, respectively. The stable fixed point $(V_0,W_0)=(0,0)$ (blue dot) lies to the right of the minimum of the $V-$nullcline and is surrounded by an unstable limit cycle (red) and a stable limit cycle (blue).  \marius{Throughout the remainder of this paper, we fix the parameter values of all the FHN neurons such that each resides within the bi-stable regime}; i.e., we set $a=-0.05$, $b=1.0$, and $c=2.0$. 

\subsection{ISR in a single FHN neuron}
\marius{It is worth pointing out that when noise is introduced into the deterministic FHN neuron (i.e., $\sigma\neq0$), the stable fixed point and stable limit cycle become meta-stable in the stochastic FHN neuron. Thus, the bi-stable regime/interval consisting of a strictly stable fixed point and stable limit cycle becomes a bi-meta-stable regime/interval consisting of a meta-stable fixed point and a meta-stable limit cycle.} 

\begin{figure}[htp!]
\centering
\includegraphics[width=\columnwidth]{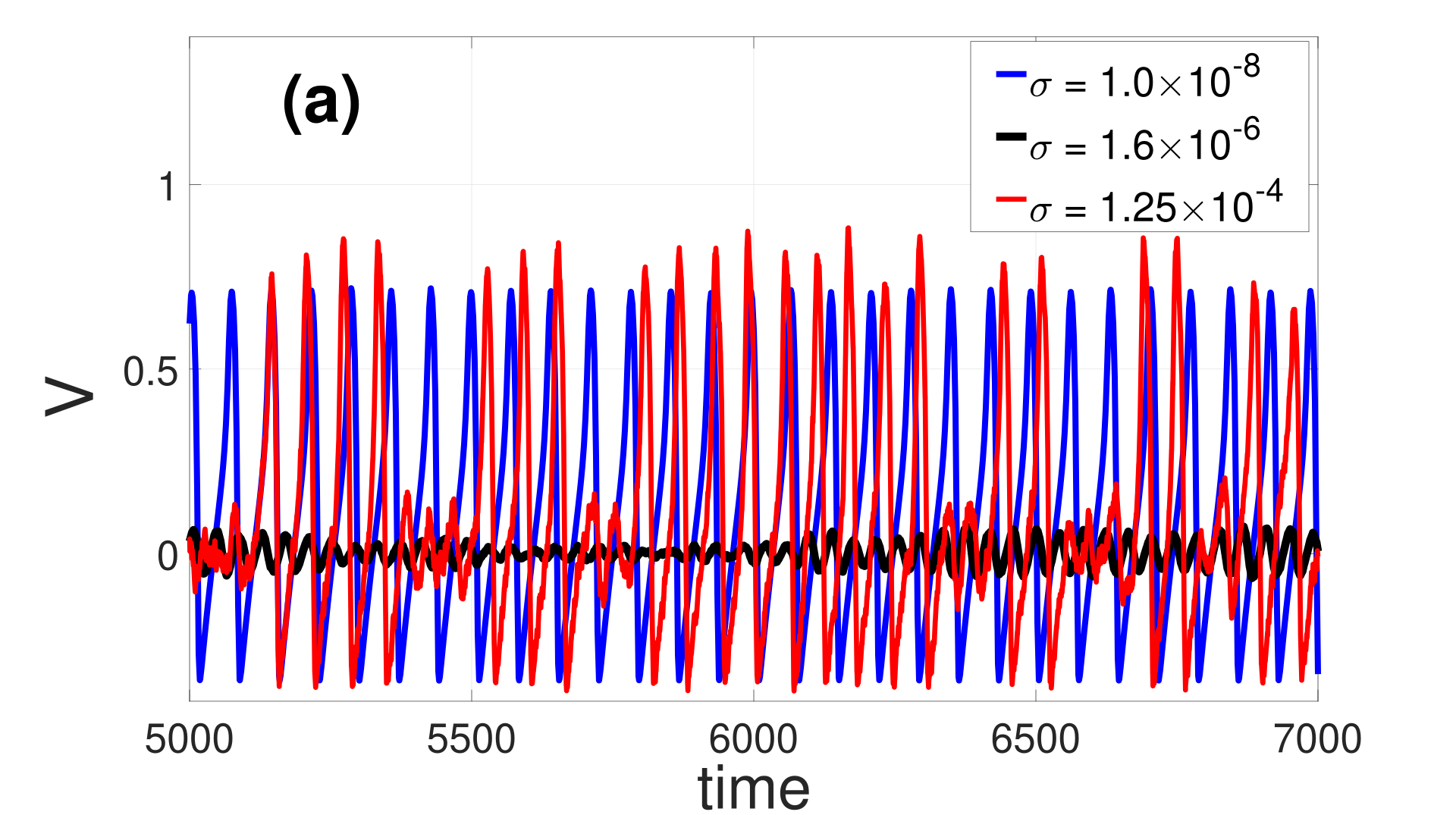}
\includegraphics[width=\columnwidth]{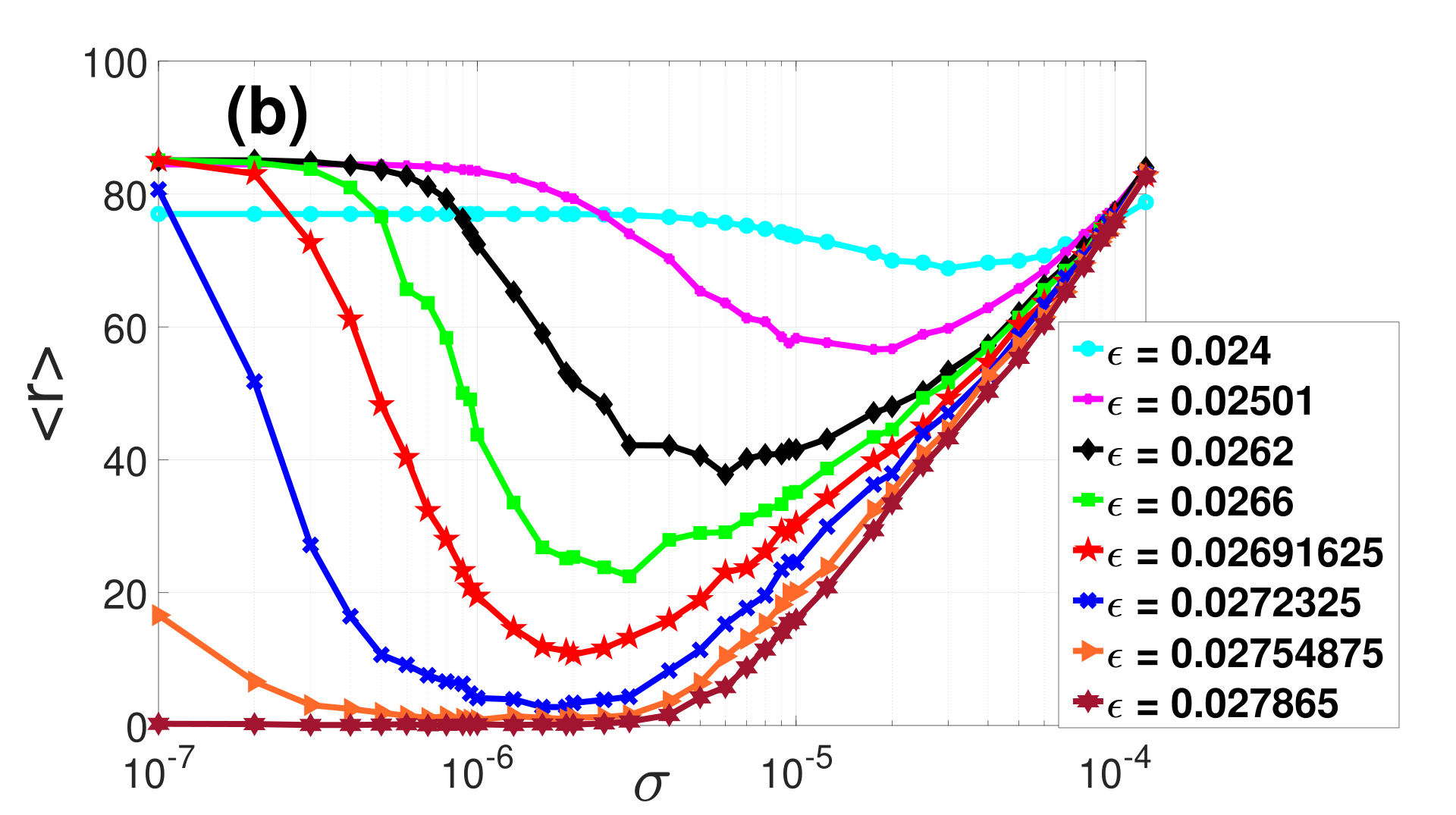}
\caption{\textbf{(a)} Time series of a single FHN neuron of the membrane potential $V$ with $\varepsilon =0.0266$ for three different values of the noise amplitude $\sigma$ indicated. 
\textbf{(b)} ISR in a single FHN neuron for moderate noise level is characterized by the non-monotonicity of mean firing rate \marius{$\overline{r}$} curves. The dependence of mean firing rate \marius{$\overline{r}$} on noise amplitude $\sigma$ for different timescale separation parameters $\varepsilon $.
$a=-0.05$, $b=1.0$, $c= 2.0$.}
\label{fig:2}
\end{figure}

We illustrate in Fig.~\ref{fig:2} how different noise intensities impact the spiking behavior of a \marius{single} FHN neuron of Eq.~\eqref{eq:1}. Initial conditions lie in the basin of attraction of the stable limit cycle (i.e., $V_i=1.0$ and $W_i=0.2$), and we examine a range of timescale parameter values $\varepsilon $ within the bi-stability interval $[0.025, 0.027865)$. 

When $\varepsilon =0.0266$,  the time series of the membrane potential $V$ in Fig.~\ref{fig:2}\textbf{(a)} shows that there is an intermediate noise intensity that \marius{inhibits} the spiking activity, see the black time series with $\sigma =1.6\times10^{-6}$. In Fig.~\ref{fig:2}\textbf{(b)}, the behavior of the mean firing rate \marius{$\overline{r}$} is shown with respect to varying both the noise intensity $\sigma$ and the timescale parameter $\varepsilon $. The \textit{non-monotonic} behavior of the mean firing rate \marius{$\overline{r}$}  as $\sigma$ and  $\varepsilon $ vary is characteristic of ISR. 

A stronger ISR effect is associated with a deeper minimum in the non-monotonic \marius{$\overline{r}$} curve. Inspecting Fig.~\ref{fig:2}\textbf{(b)}, it becomes evident that the closer the timescale parameter $\varepsilon $ is to its Hopf bifurcation value $\varepsilon _\text{HB}=-a/c=0.025$, the weaker is the ISR effect, see, e.g., the magenta curve with $\varepsilon =0.02501$. As $\varepsilon $ increases within the \marius{bi-meta-stable interval}, the non-monotonic \marius{$\overline{r}$} curve becomes deeper, indicating a stronger ISR effect; see, e.g., the purple curve in Fig.~\ref{fig:2}\textbf{(b)} with $\varepsilon =0.0272325$.

We also notice that when $\varepsilon $ is outside the \marius{bi-meta-stable interval} $[0.025, 0.027865)$, e.g., for $\varepsilon =0.0278650$, the effect of ISR disappears with the disappearance of the \marius{bi-metastability} between the fixed point $(V_0,W_0)=(0,0)$ and the limit cycle. For $\varepsilon \geq0.0278650$ (see Fig.~\ref{fig:2}\textbf{(b)}), only the stable fixed point $(V_0,W_0)=(0,0)$ remains, as the stable and unstable limit cycles have collided and annihilated each other. Consequently, increasing the noise intensity $\sigma$ results in a monotonic, rather than non-monotonic, increase in the mean firing rate \marius{$\overline{r}$}.

The strengthening of the ISR effect with increasing $\varepsilon \in[0.025, 0.027865)$ can be explained in terms of the basin of attractions. Near the Hopf bifurcation threshold $\varepsilon _\text{HB}=0.025$, the basin of attraction for the fixed point $(V_0,W_0)=(0,0)$ is significantly smaller in comparison to that of the stable limit cycle. Consequently, trajectories tend to swiftly depart from the basin of the fixed point while lingering longer within the basin of the limit cycle. This yields, as the noise intensity increases, a shallow \marius{$\overline{r}$} curve, as depicted by the magenta curve in Fig.~\ref{fig:2}\textbf{(b)} for $\varepsilon =0.02501$.

As $\varepsilon \in[0.025, 0.027865)$ increases, the basin of attraction of the stable fixed point expands while that of the stable limit cycle contracts (eventually vanishing at $\varepsilon =0.027865$). Consequently, the exit times from the basin of the fixed point lengthen compared to those from the limit cycle, leading to non-monotonic \marius{$\overline{r}$} curves characterized by deeper minima as $\varepsilon \in[0.025, 0.027865)$ increases.

\marius{Furthermore, in Fig.~\ref{fig:2}\textbf{(b)}, one can observe the occurrence of a very weak ISR when $\varepsilon(=0.024)$ is outside (but also very close to the sub-critical Hopf bifurcation threshold at $\varepsilon=0.025$) the bi-stability interval. This occurrence is explained by the fact that noise can temporally stabilize the unstable fixed point which is at the edge of the Hopf bifurcation threshold. Hence, trajectories can spend a slightly longer time in the basin of attraction of this fixed point, inducing the weak ISR seen in Fig.~\ref{fig:2}\textbf{(b)} at $\varepsilon=0.024$. See Ref. \cite{bavcic2020two} for more details on the mechanism of ISR induced via noise-enhanced stability of a fixed point. We emphasize that in the current study, the focus is on the mechanism of ISR only within the \marius{bi-meta-stable interval}.}

\subsection{ISR in the adaptive network}

\subsubsection{Effect of $\varepsilon $ on ISR with STDP only}
\begin{figure}
\centering
\includegraphics[width=\columnwidth]{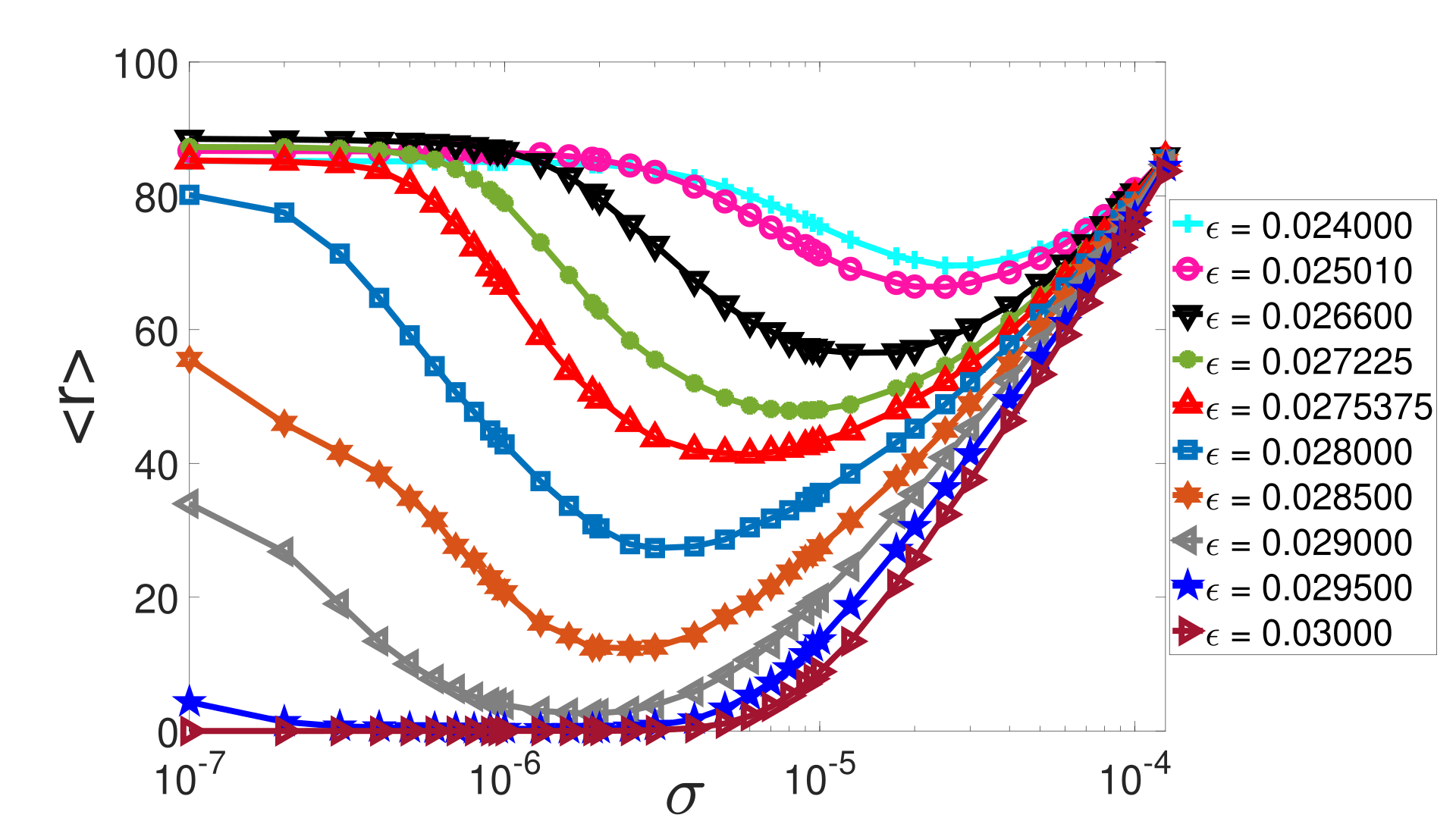}
\caption{Collective firing behavior of a SW network with STDP (and without HSP) for different timescale parameter values $\varepsilon $ and noise amplitudes $\sigma$. The emergence of ISR in this network is characterized by the non-monotonicity of the mean firing rate $\langle r \rangle$ curves that occur at intermediate noise levels. Stronger ISR occurs at the largest value of  $\varepsilon \in[0.024,0.029]$. 
$a=-0.05$, $b=1.0$, $c= 2.0$, $V_\text{syn}=2.0$, $V_\text{shp}=0.05$, $\beta=0.25$, $\langle k \rangle=4$, $\tau_a=\tau_b=2.0$, $B=0.5$, $P=5.0\times10^{-6}$, $A=B/P$, $F=0.0$, $N=70$.}
\label{fig:3}
\end{figure}

Since the timescale parameter $\varepsilon $ controls the size of the basin of attraction of both the stable fixed point and the limit cycle in the isolated neuron, it is natural to investigate how $\varepsilon $  affects the degree of ISR in an adaptive network. 
To do this, we computed the mean firing rate $\langle r \rangle$ in the STDP-driven SW network without HSP (i.e., $F=0$ Hz) while varying $\varepsilon $, see  Fig.~\ref{fig:3}. \marius{It is worth pointing out that, similar to how ISR can occur outside the \marius{bi-meta-stable interval} in an isolated neuron due to noise-enhanced stabilization of the fixed point \cite{bavcic2020two}, in Fig.~\ref{fig:3} this effect adds up and amplifies in a network with many neurons. Consequently, the range of $\varepsilon$ (i.e., $\varepsilon\in[0.024,0.029]$) where ISR can still be observed in the network slightly extends beyond the \marius{bi-meta-stable interval} (i.e., $\varepsilon\in[0.025, 0.027865)$) of individual neurons, and even further than the extended interval (i.e., $\varepsilon\in[0.024, 0.027865)$ due to noise-enhanced stability) where ISR is observed in the isolated neuron.}

The dynamic behavior of neurons {\it in vivo} has to be characterized as a collective phenomenon rather than in isolation, and our simulation describes indeed the setting of a neural network. Thus, our results (see Fig.~\ref{fig:3}) strongly suggest that ISR should be more pronounced and more readily observed in experiments involving networks of neurons when compared to isolated neurons~\cite{Buchin2016PlosCB}. Our observations for these simulations are qualitatively similar to Fig.~\ref{fig:2}\textbf{(b)}, i.e., the minimum mean firing rate $\langle r \rangle_\text{min}$ decreases as $\varepsilon $ increases, thus indicating an enhanced inhibition of the spiking activity by ISR. 

\begin{figure}
\centering
\includegraphics[width=\columnwidth]{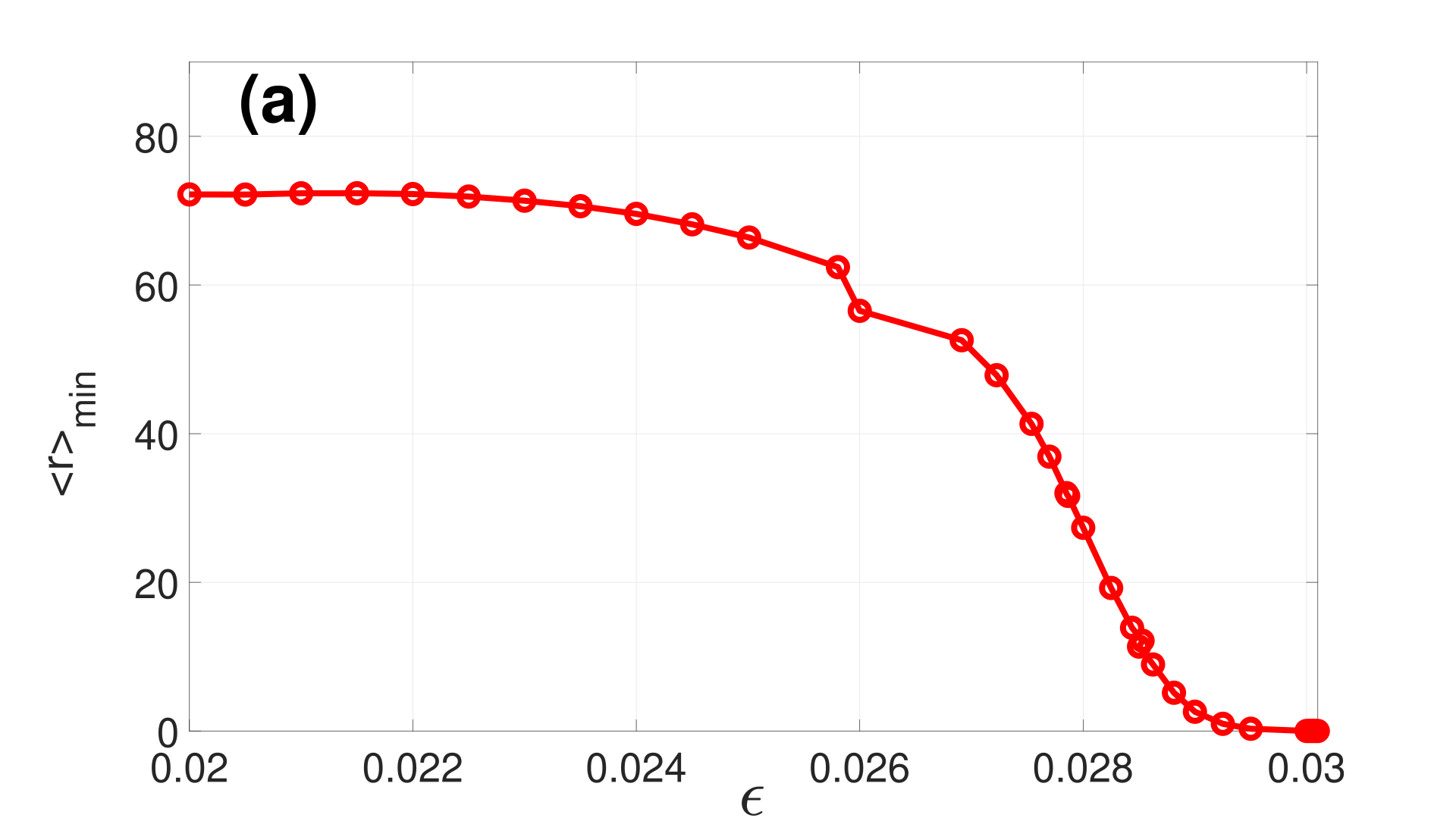}
\includegraphics[width=\columnwidth]{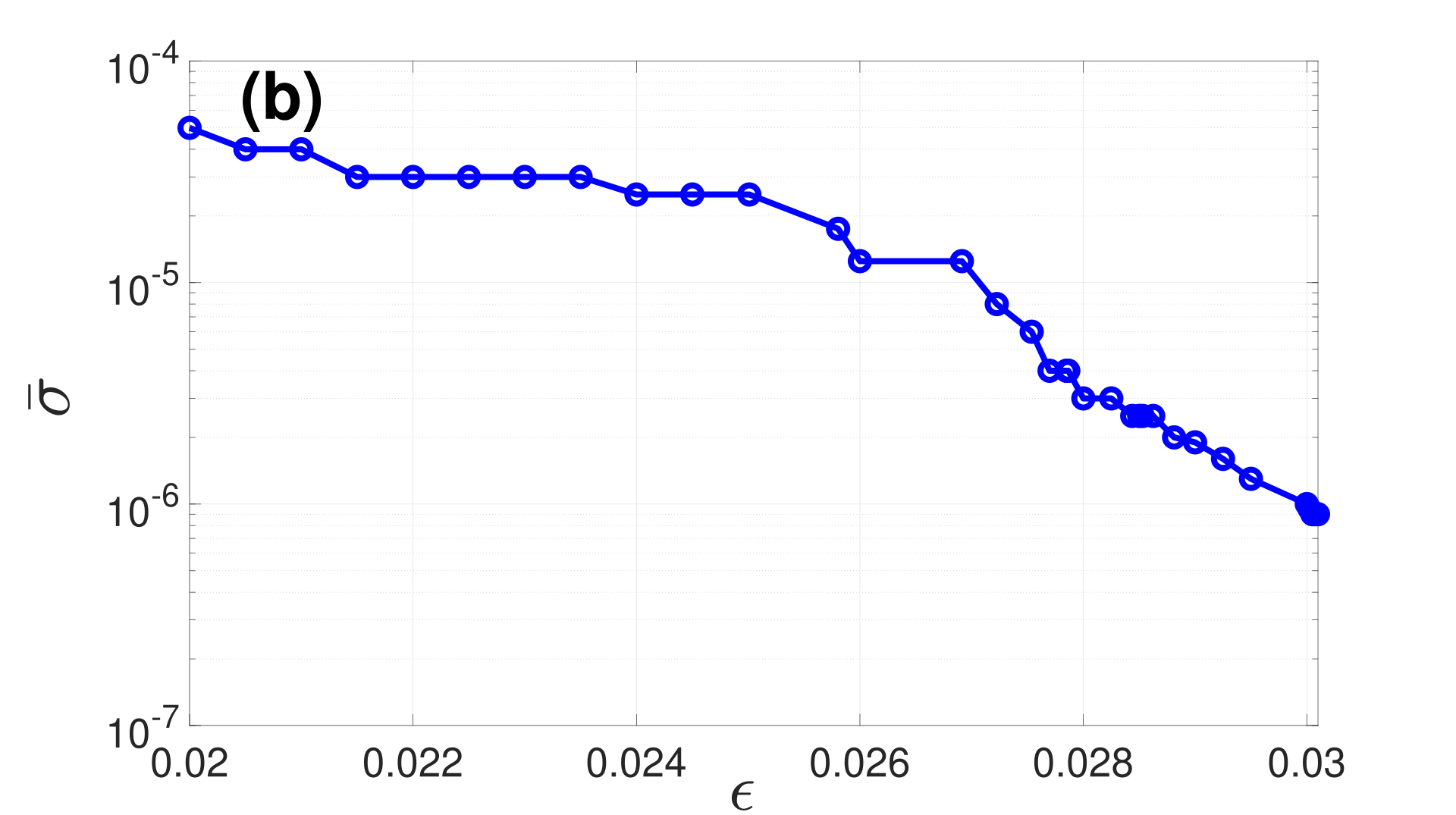}
\includegraphics[width=\columnwidth]{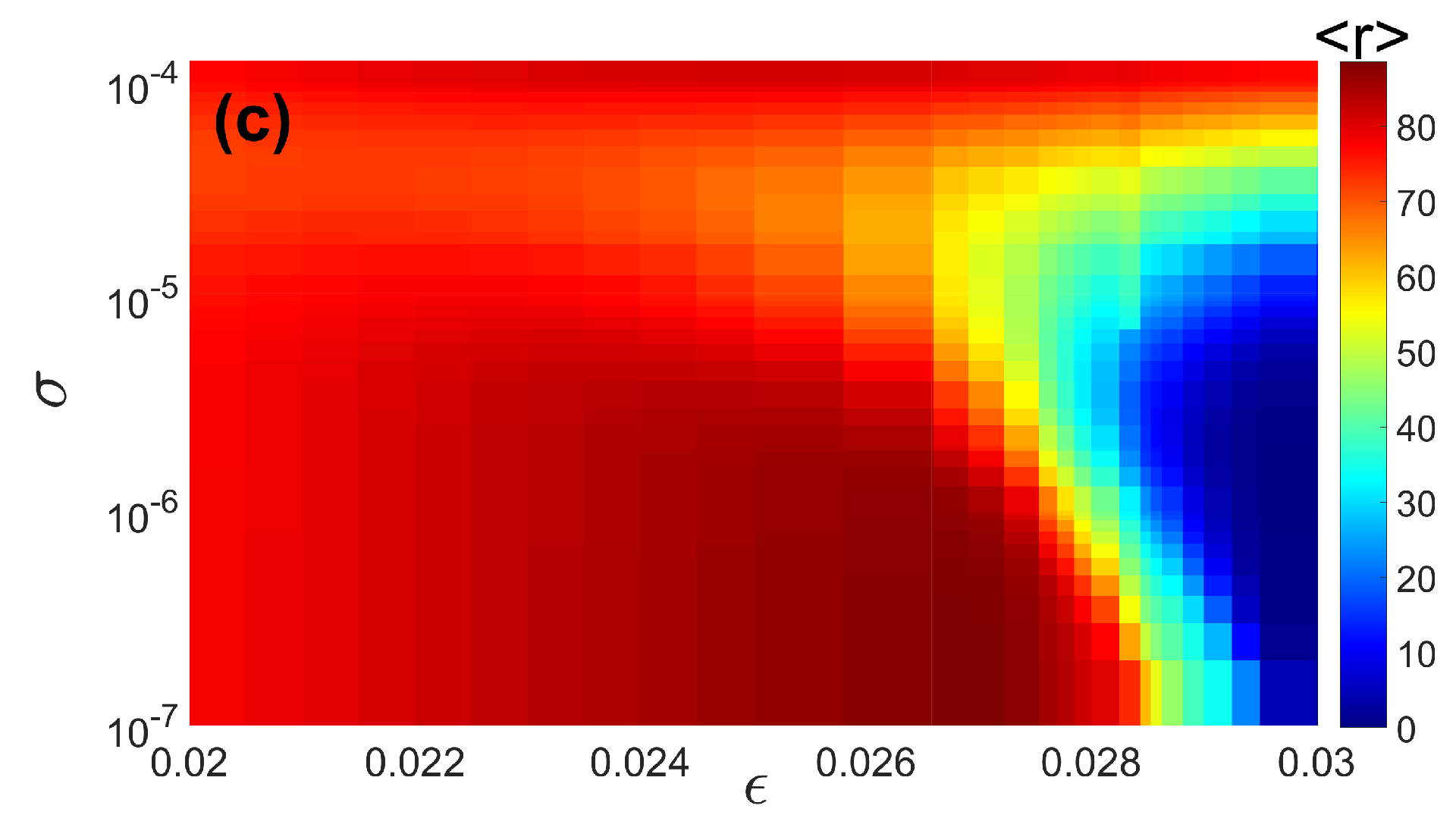}
\caption{Absence and appearance of ISR in an STDP-driven small-world network without HSP. \textbf{(a)} The
minima of the mean firing rate $\langle r \rangle_\text{min}$ vs. $\varepsilon $. \textbf{(b)} Corresponding noise intensity
$\overline{\sigma}$ to $\langle r \rangle_\text{min}$ response to
the timescale parameter $\varepsilon $. 
\textbf{(c)} Mean firing rate $\langle r \rangle$ in dependence of $\varepsilon $ and $\sigma$ plane displays ISR for $\varepsilon \in[0.024,0.029]$. Parameters are:
$a=-0.05$, $b=1.0$, $c= 2.0$, $V_\text{syn}=2.0$, $V_\text{shp}=0.05$, $\beta=0.25$, $\langle k \rangle=4$, $\tau_a=\tau_b=2.0$, $B=0.5$, $P=5.0\times10^{-6}$, $F=0.0$, $A=B/P$, $N=70$.}
\label{fig:4}
\end{figure}

However, this effect gets stronger when $\varepsilon \geq 0.027$ and $\langle r \rangle_\text{min}$ becomes very low around $\varepsilon \approx 0.028$ --- close to the upper boundary of the bi-stability region (see Fig.~\ref{fig:4} \textbf{(a)}). This behavior can be explained by the rapid coalescence of the stable and the unstable limit cycles at the fold bifurcation as $\varepsilon $ increases, see Fig.~\ref{fig:1}. The ensuing rapid decrease in the size of the basin of attraction of the limit cycle also induces a rapid decrease in the optimal noise intensity $\overline{\sigma}$ required to achieve $\langle r \rangle_\text{min}$ (see, e.g., Fig.~\ref{fig:4}\textbf{(b)}). The plot of $\langle r \rangle$ in the $(\varepsilon ,\sigma)$-plane (Fig.~\ref{fig:4}\textbf{(c)}) clearly reveals this effect at the boundary, in agreement with Fig.~\ref{fig:4}\textbf{(a)} and \textbf{(b)}.

%%%%%%%%%%%%%%%%%%%

\begin{figure}[h]
\centering
\includegraphics[width=\columnwidth]{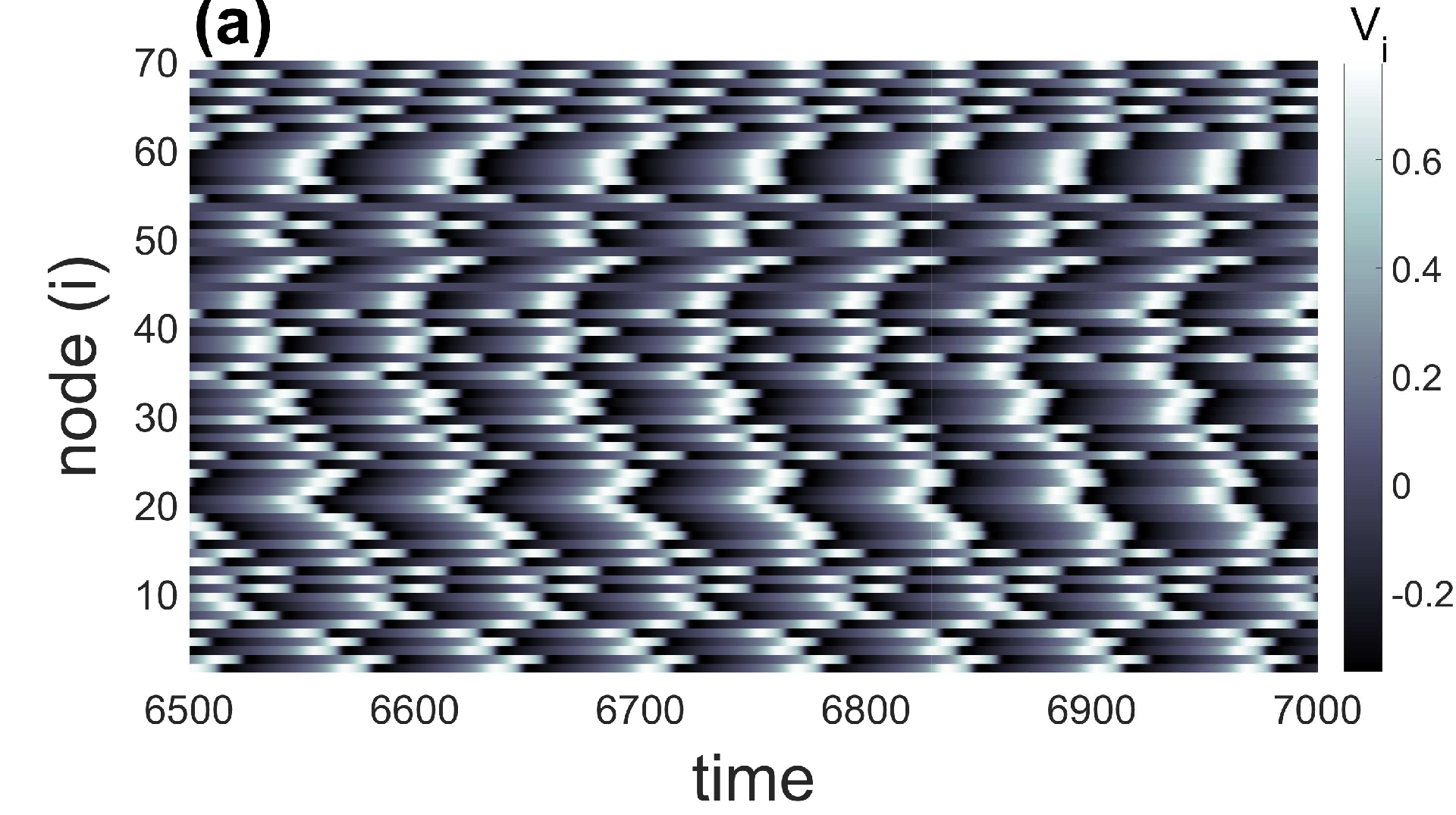}
\includegraphics[width=\columnwidth]{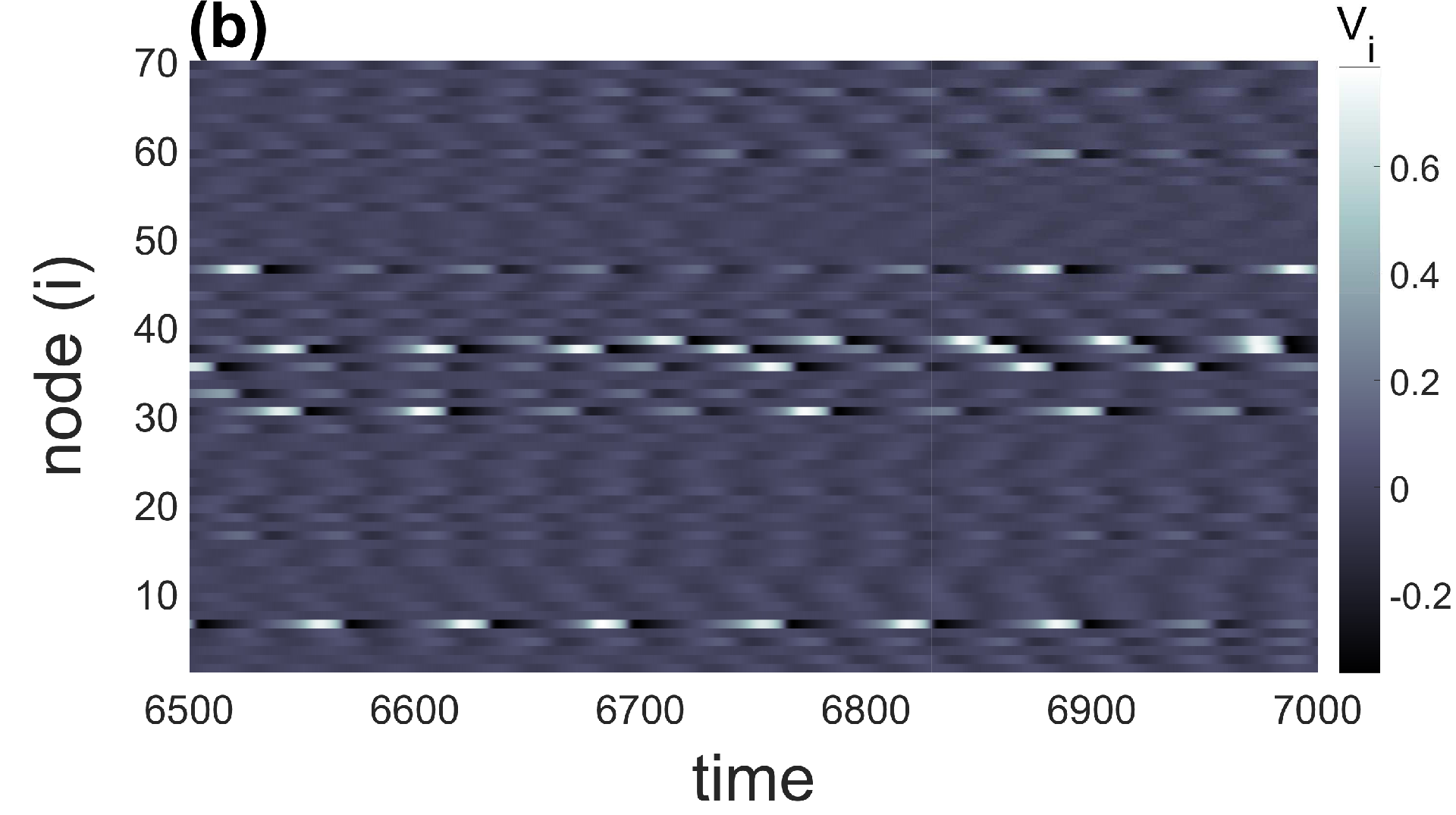}
\includegraphics[width=\columnwidth]{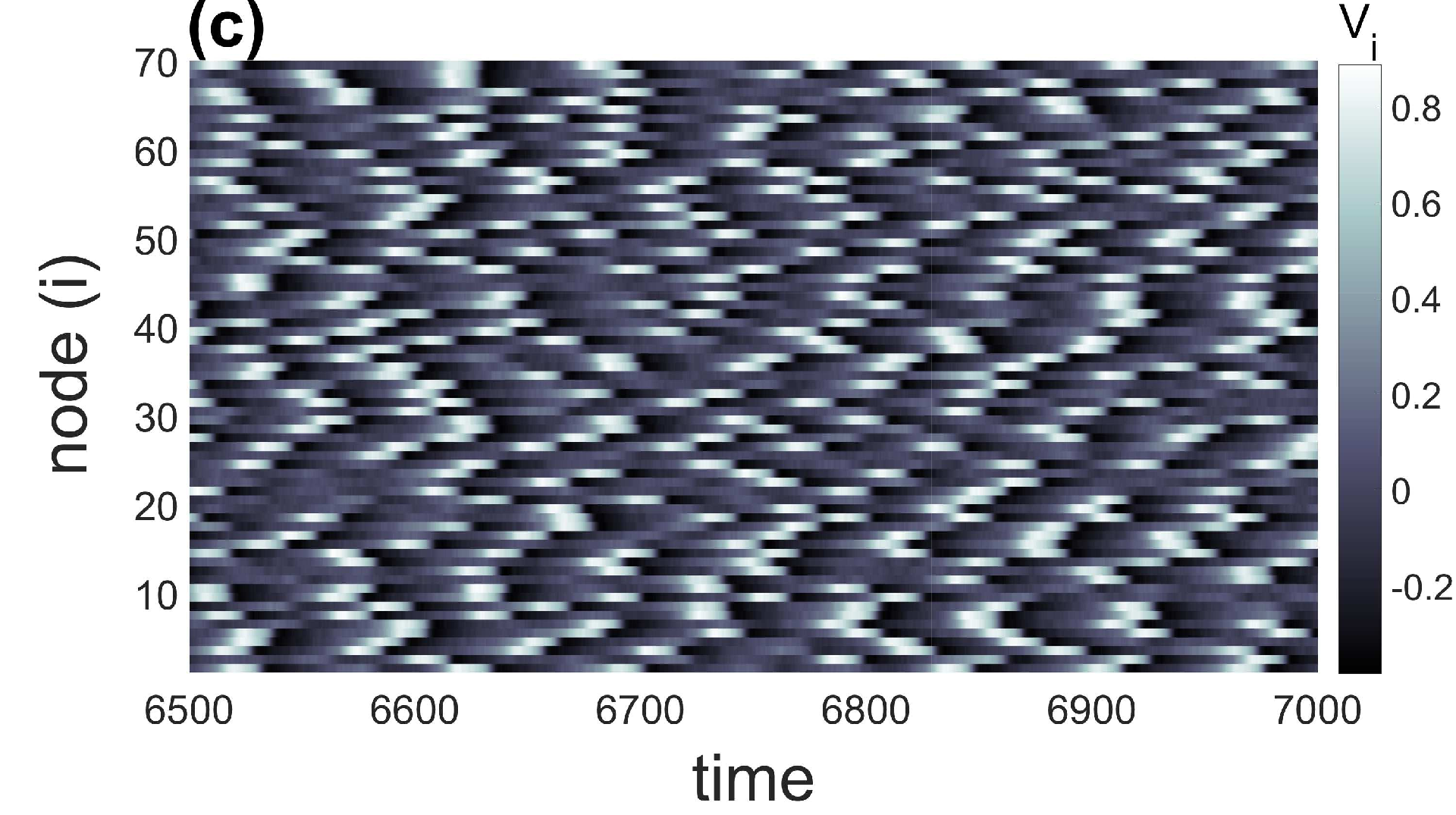}
\caption{Dynamic behavior of neurons in the STDP-driven SW network without HSP at three different noise amplitudes for $\varepsilon =0.0285$: 
\textbf{(a)} $\sigma= 1.0\times10^{-8}$;
\textbf{(b)}  $\sigma= 2.5\times10^{-6}$;  
\textbf{(c)}  $\sigma=1.25\times10^{-4}$.
$a=-0.05$, $b=1.0$, $c= 2.0$, $V_\text{syn}=2.0$, $V_\text{shp}=0.05$, $\beta=0.25$, $\langle k \rangle=4$, $\tau_a=\tau_b=2.0$, $B=0.5$, $P=5.0\times10^{-6}$, $A=B/P$, $F=0.0$, $N=70$.}
\label{fig:5}
\end{figure}

\marius{Figure~\ref{fig:5} shows examples of the behavior} of the network for $\varepsilon =0.0285$ with three different noise intensities. The neurons in the adaptive networks spike frequently for small and large noises. We observe a quasi-total inhibition of the spiking activity due to ISR for the intermediate noise.
%%%%%%%%%%%%%%%%%%%

\subsubsection{Effect of HSP on ISR}
%%%%%%%%%%%%%%%%%%%
 
\begin{figure}
\centering
\includegraphics[width=\columnwidth,height=50.0mm]{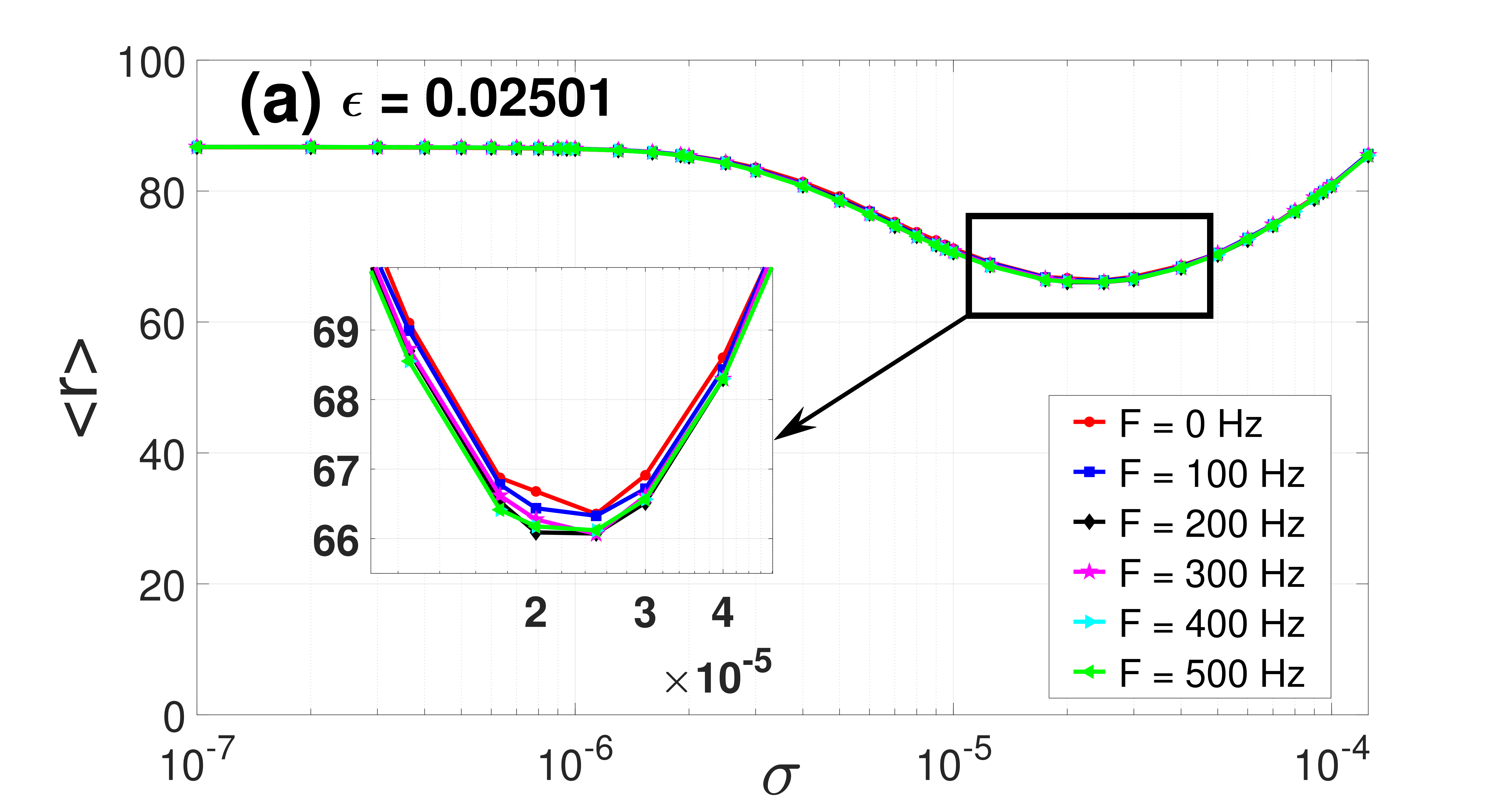}
\includegraphics[width=\columnwidth,height=50.0mm]{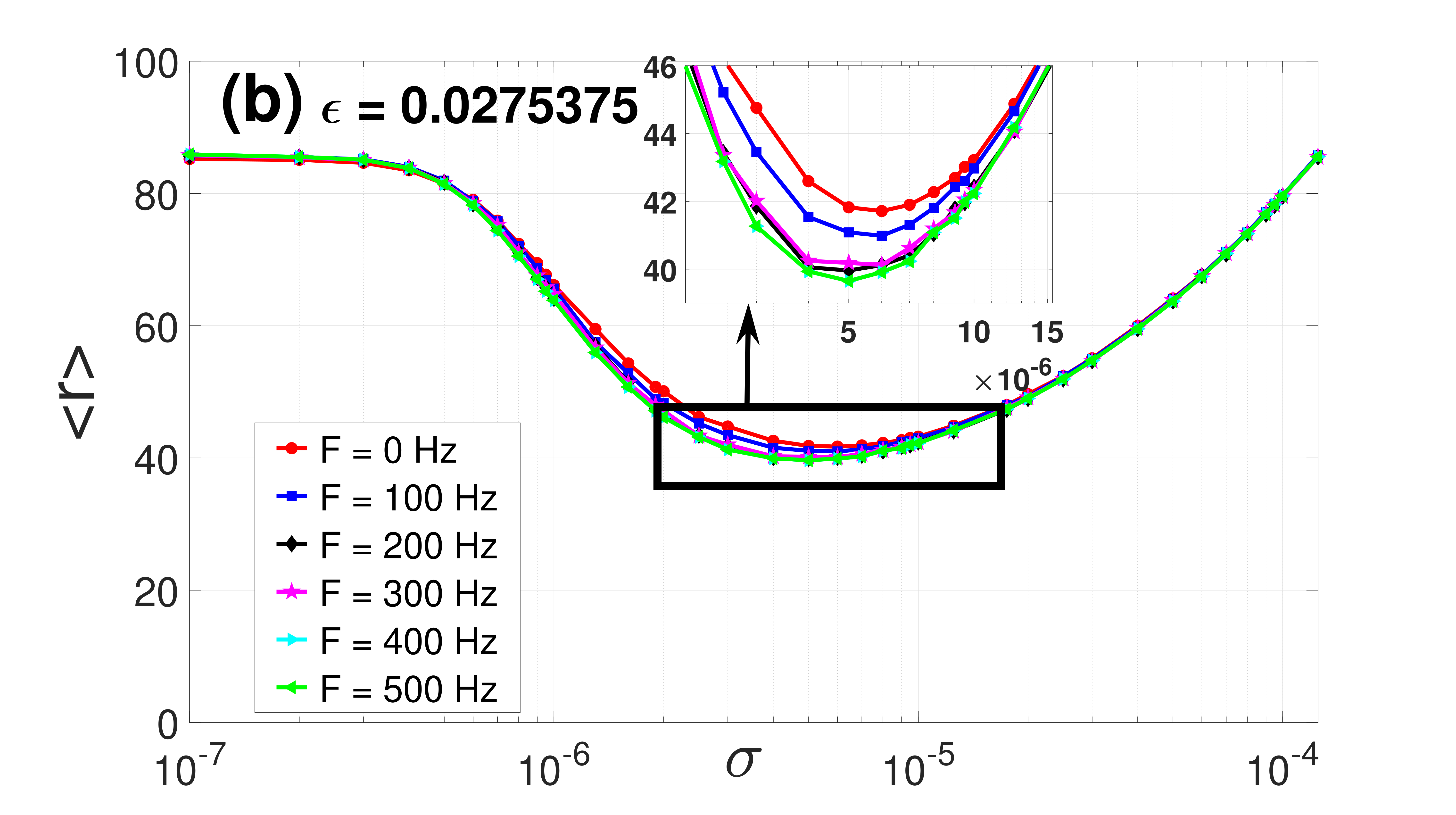}
\includegraphics[width=\columnwidth,height=50.0mm]{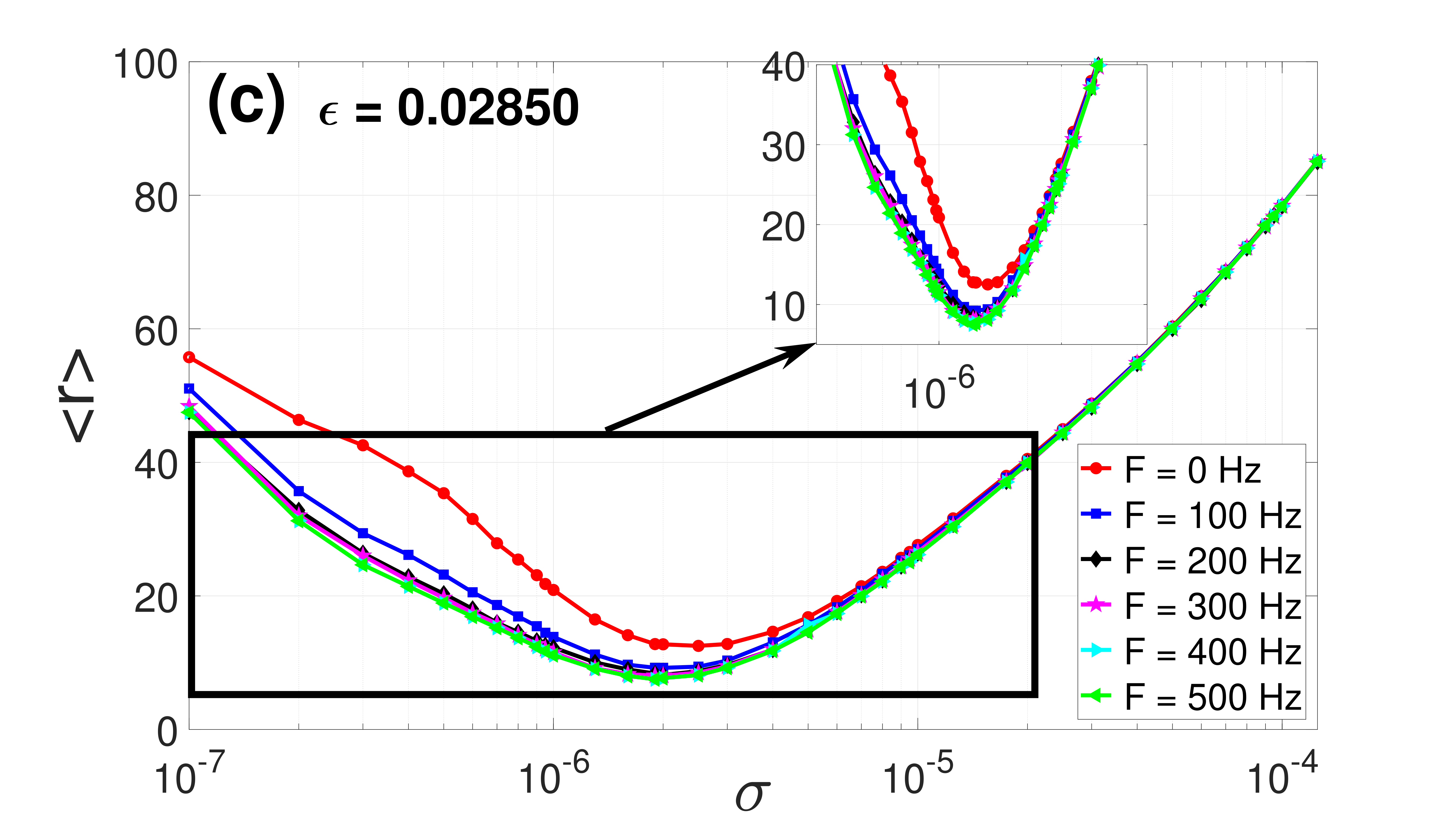}
\includegraphics[width=\columnwidth,height=50.0mm]{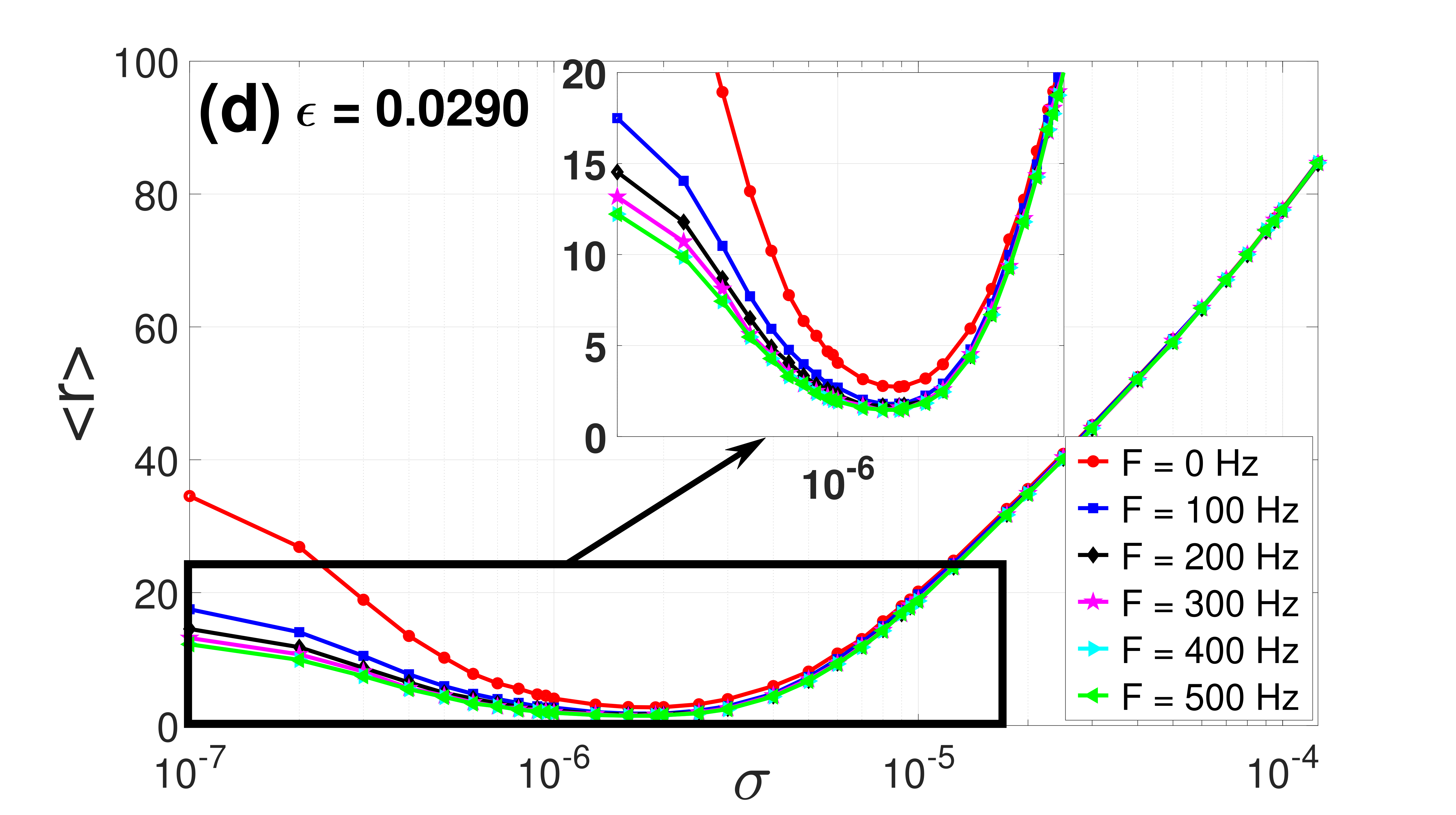}
\caption{Effect of HSP (controlled by $F$) on collective firing behavior of a SW network at four different values of $\varepsilon $. Higher $F$ lowers the $\langle r \rangle$ curves, thereby enhancing ISR. However, this effect is significant for larger values of $\varepsilon $ and weak noise amplitudes. $a=-0.05$, $b=1.0$, $c= 2.0$, $V_\text{syn}=2.0$, $V_\text{shp}=0.05$, $\beta=0.25$, $\langle k \rangle=4$, $\tau_a=\tau_b=2.0$, $B=0.5$, $P=5.0\times10^{-6}$,  $A=B/P$, $N=70$.}
\label{fig:6}
\end{figure}

We study the effects of only HSP on ISR, see Fig.~\ref{fig:6}. Recall that the HSP algorithm rewires the synaptic connections between neurons while preserving the small-worldness of the network topology (see Sec.~\ref{Sec. III}). The rewiring frequency $F$ determines how fast the SW network reshapes its synaptic connections. We investigate a wide \marius{range of $F$ from} $0$ Hz to $500$ Hz, which covers the limiting regimes where the network is static (i.e., rewires with probability zero) up to where the rewiring of synapses between two distant neurons occurs with probability close to unity (i.e., rewires at each time step and hence, changes the topology very quickly). Fig.~\ref{fig:6} illustrates how $F$ affects the mean \marius{firing} frequency $\langle r \rangle$ and ISR, as we vary the rewiring frequency $F$ while keeping all other parameters fixed. 

The overall trend is that faster rewiring implies a lower ISR curve; however, this behavior may be more or less pronounced as we vary $\varepsilon $. 
Specifically, when $\varepsilon $ is close to the lower boundary of the bi-stability region (see Fig.~\ref{fig:6}\textbf{(a)} where $\varepsilon =0.02501$), the enhancement of ISR by increasing $F$ is inconspicuous compared with when $\varepsilon $ is \marius{chosen further into} the \marius{bi-meta-stable interval}. 
With larger values of $\varepsilon $ as in Figs.~\ref{fig:6}\textbf{(b)}-\textbf{(d)}, the enhancement of ISR becomes stronger for higher rewiring frequency $F$. Furthermore, we observe (especially in Fig.~\ref{fig:6}\textbf{(b)}) that the optimal noise intensity, i.e., where the mean firing rate is most inhibited and ISR is most pronounced, is slightly shifted to the left towards smaller $\sigma$ values. 
Thus, we observe that HSP results in \emph{two effects}: (i) larger $F$ lowers the ISR curves, and (ii) larger $F$ slightly shifts the minima of the ISR curves to the left, making them occur at slightly smaller values of the noise intensity $\sigma$.

Conceptually, we may explain these effects as follows.
First, note that neurons in the network have independent noise sources and different initial conditions. Therefore, they can spike independently of each other and at different times. Second, fast rewiring (i.e., larger $F$) makes it even more difficult for the neurons to synchronize their spiking activity, as neurons constantly swap neighbors via HSP. When $F$ is large, connecting neurons do not have the time to synchronize their different spiking times before they become disconnected again via HSP. Hence, neurons in the network spiking at different rates would quickly and repeatedly connect and disconnect from each other as time evolves. The overall consequence of the quick connections and disconnections between neurons with varying spiking rates is the creation of a second source of synaptic noise to each neuron in the network. This second source of synaptic noise (induced and measured by $F$) combines with the synaptic noise (measured by $\sigma$) to increase the overall external stochastic forcing of the neurons involved in fast synaptic rewiring. Ultimately, this leads to a downward shift of the ISR curves, with minima occurring at lower values of the synaptic noise intensity $\sigma$ which explains effect (i).

Next, we note that larger $\varepsilon $ expands (shrinks) the basin of attraction of the fixed point (limit cycle). Thus, a smaller noise intensity is required to kick trajectories out of the smaller basin of attraction of the limit cycle. Furthermore, the additional synaptic noise source (induced by the fast rewiring of the synaptic connections) makes it even easier for trajectories to escape from the basin of attraction of the limit cycle to that of the fixed point, as it assists the small synaptic noise intensity $\sigma$ in the escape process. 
This explains effect (ii). 

\subsubsection{Effect of STDP on ISR}
\begin{figure}
\centering
\includegraphics[width=\columnwidth]{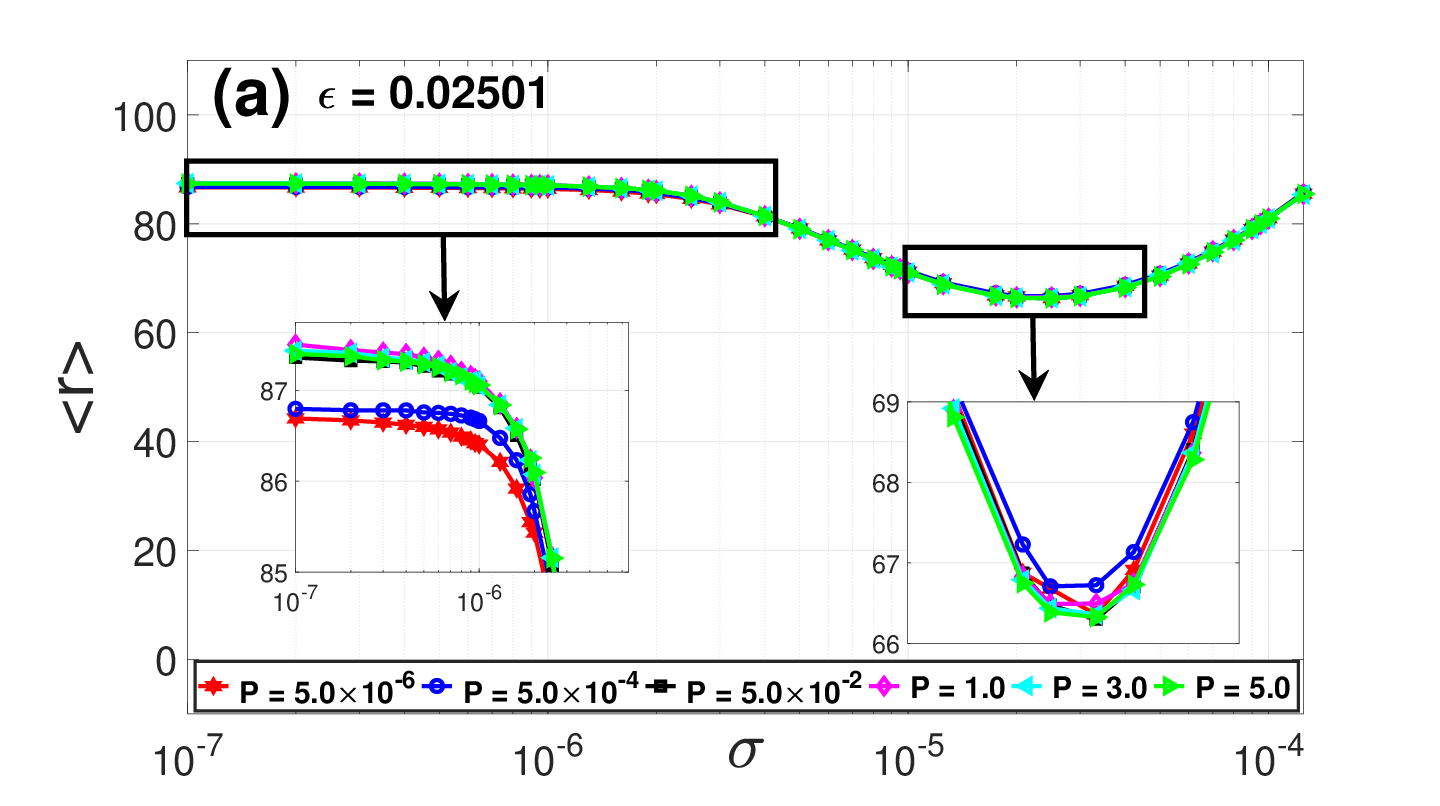}
\includegraphics[width=\columnwidth]{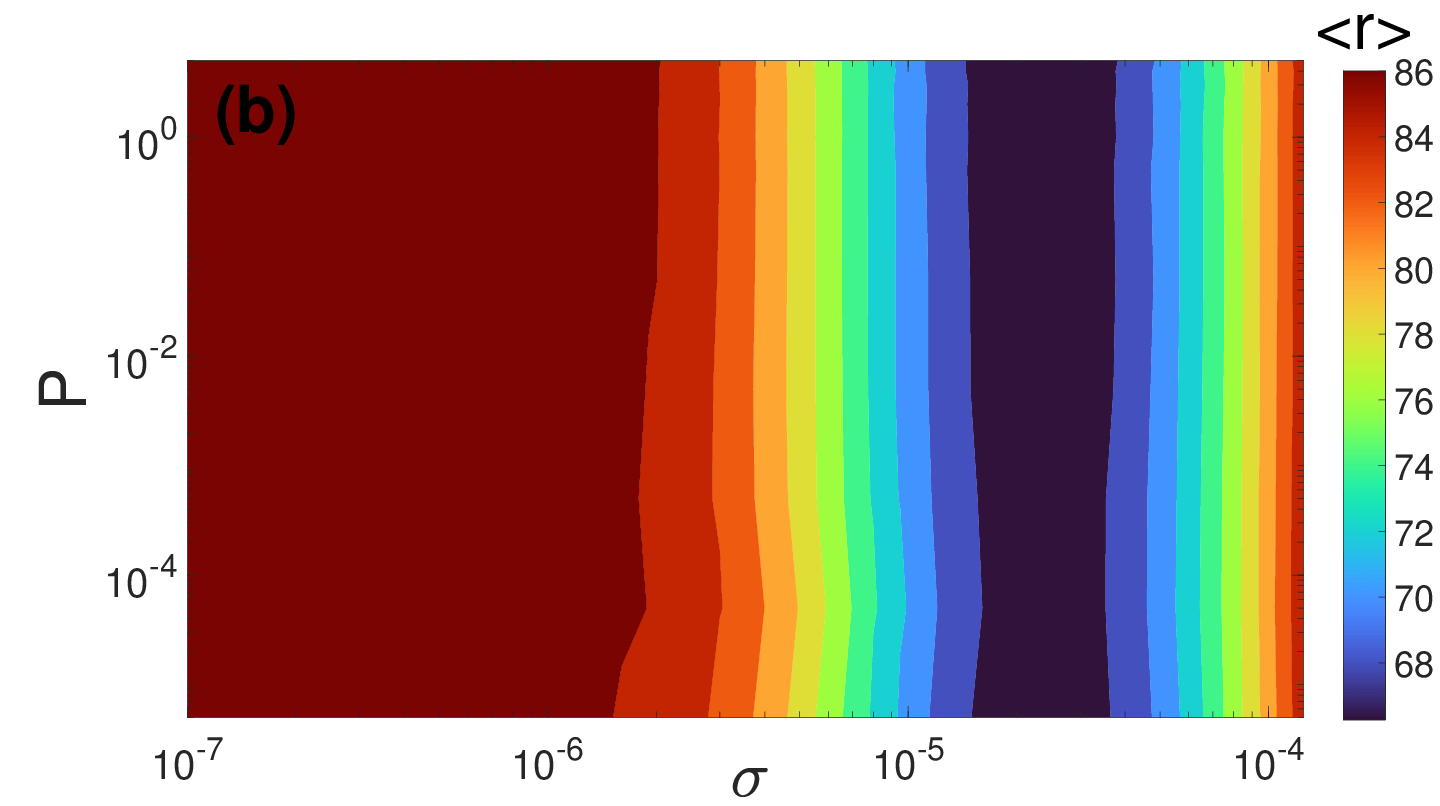}
\includegraphics[width=\columnwidth]{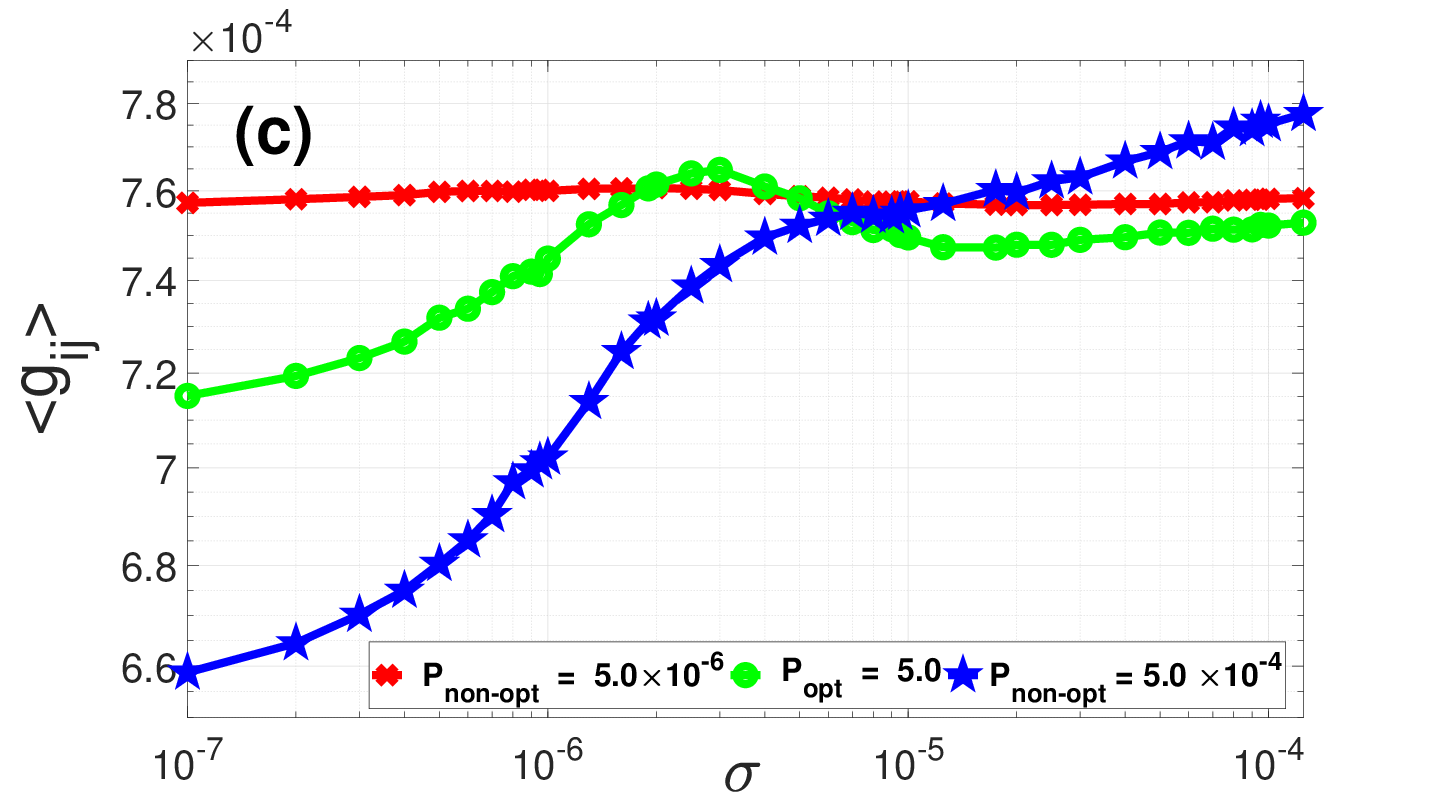}
\caption{Effect of STDP (controlled by $P$) on collective firing of the SW network at $\varepsilon =0.02501$ (panels \textbf{(a)} and \textbf{(b)}). \textbf{(c)} Associated population- and time-averaged synaptic weights of the network $\langle g_{ij} \rangle$ vs. noise $\sigma$ for two non-optimal values and the optimal value of $P$.  Parameters are: $a=-0.05$, $b=1.0$, $c= 2.0$, $V_\text{syn}=2.0$, $V_\text{shp}=0.05$, $\beta=0.25$, $\langle k \rangle=4$, $F=0.0$, $\tau_a=\tau_b=2.0$, $B=0.5$, $A=B/P$, $N=70$.}
\label{fig:7}
\end{figure}

\begin{figure}
\centering
\includegraphics[width=\columnwidth]{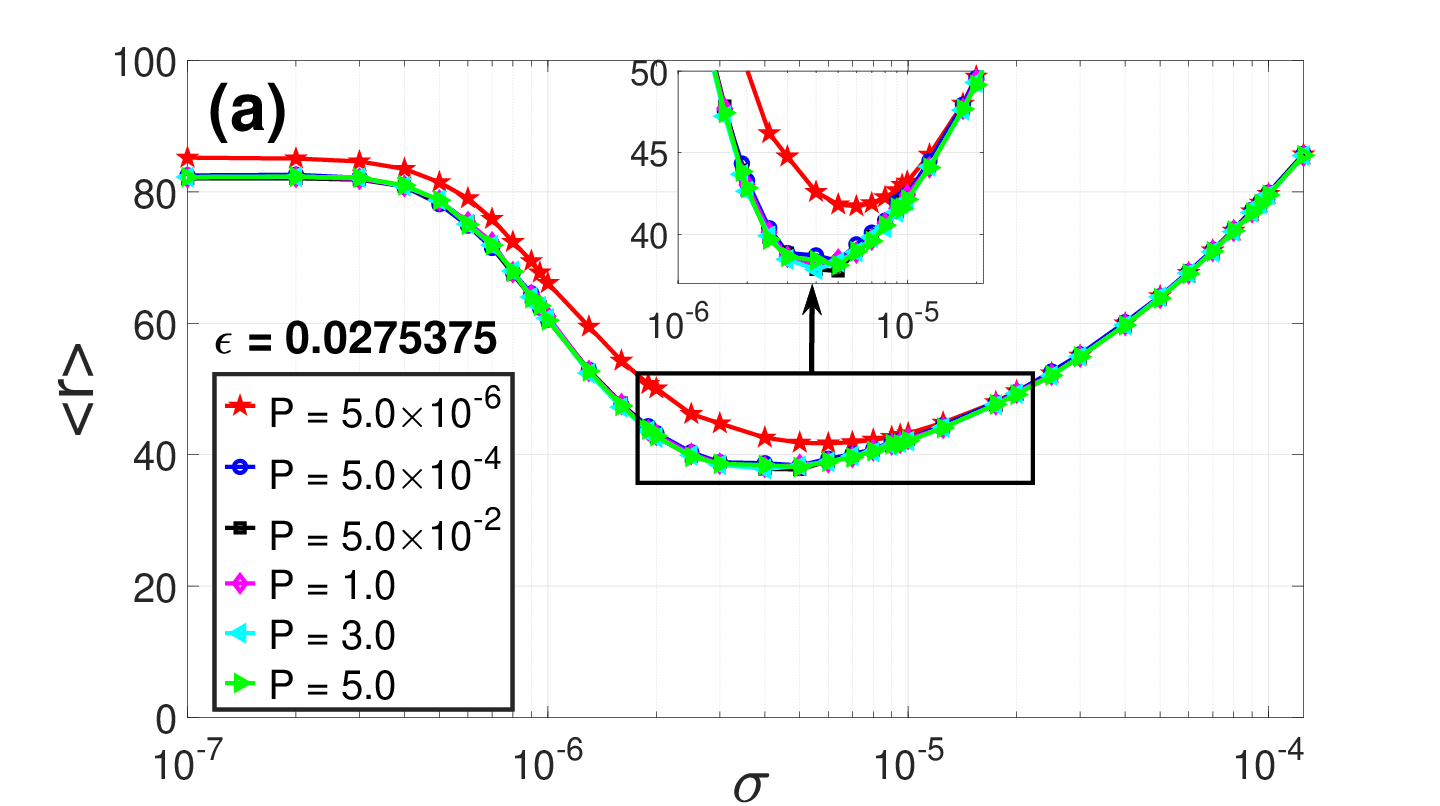}
\includegraphics[width=\columnwidth]{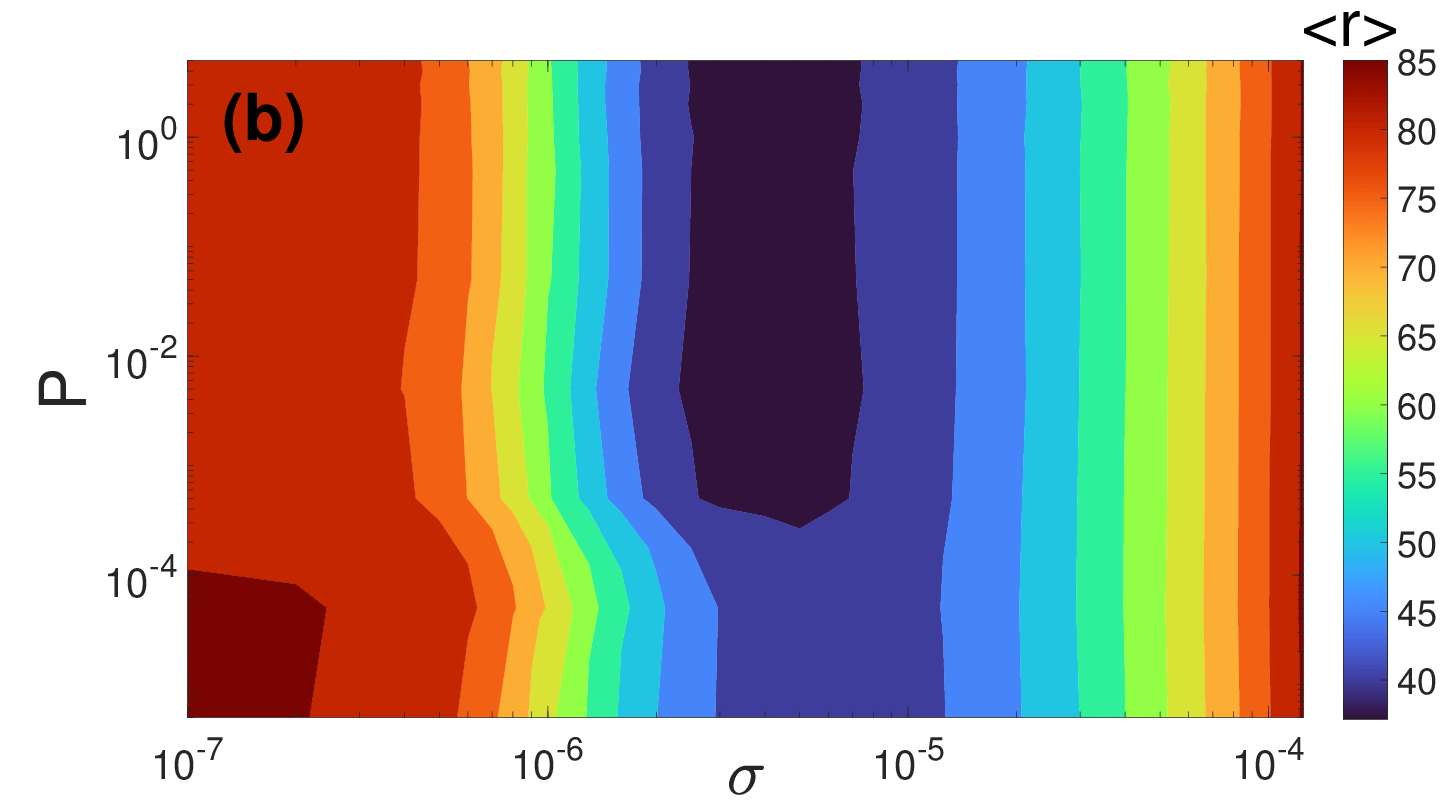}
\includegraphics[width=\columnwidth]{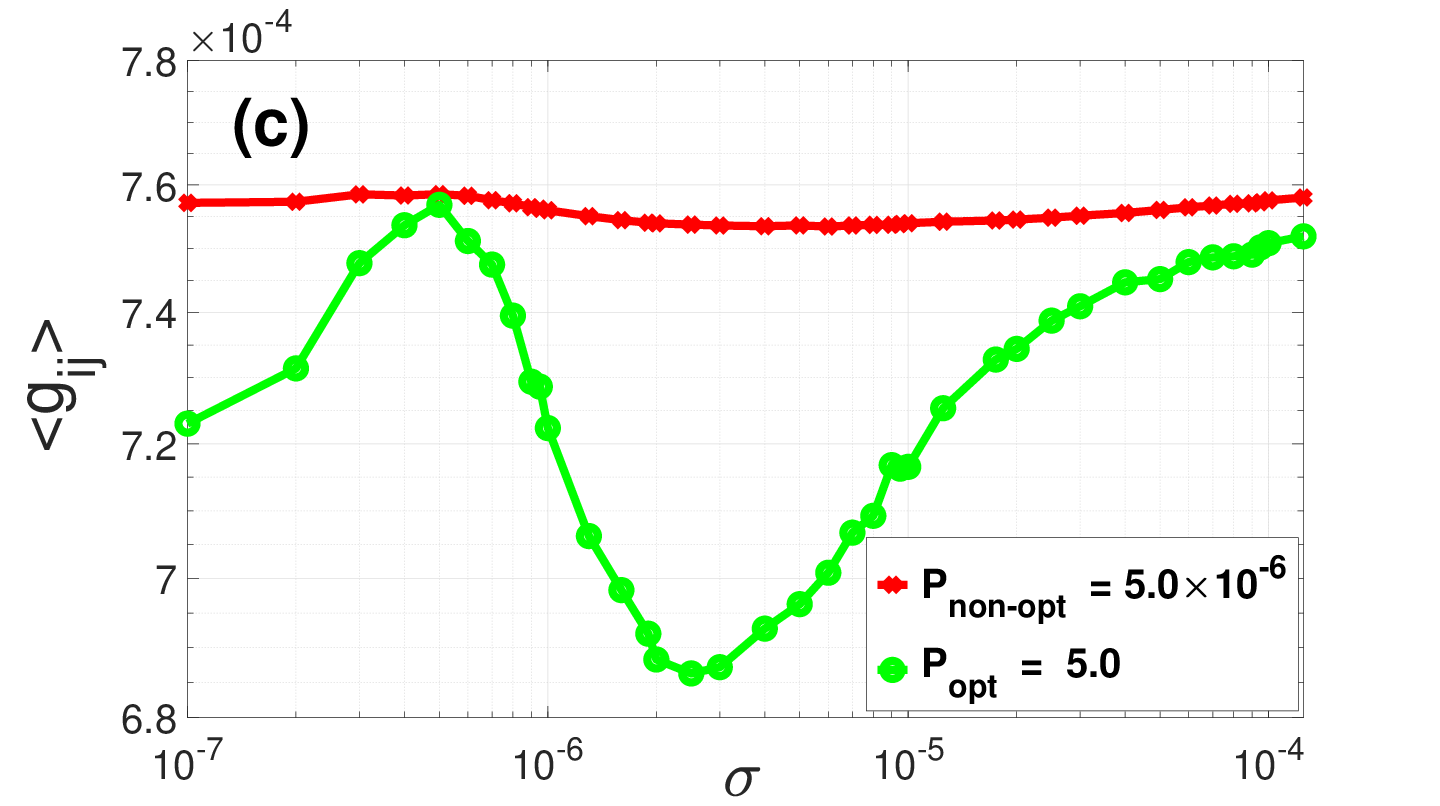}
\caption{Effect of STDP (controlled by $P$)  on collective firing behavior of the SW network at $\varepsilon =0.0275375$ in \textbf{(a)} and \textbf{(b)}. Associated population- and time-averaged synaptic weights of the network $\langle g_{ij} \rangle$ vs. noise $\sigma$ for a non-optimal value and the optimal value of $P$ in \textbf{(c)}.  $a=-0.05$, $b=1.0$, $c= 2.0$, $V_\text{syn}=2.0$, $V_\text{shp}=0.05$, $\beta=0.25$, $\langle k \rangle=4$, $F=0.0$, $\tau_a=\tau_b=2.0$, $B=0.5$, $A=B/P$, $N=70$.}
\label{fig:8}
\end{figure}

\begin{figure}
\centering
\includegraphics[width=\columnwidth]{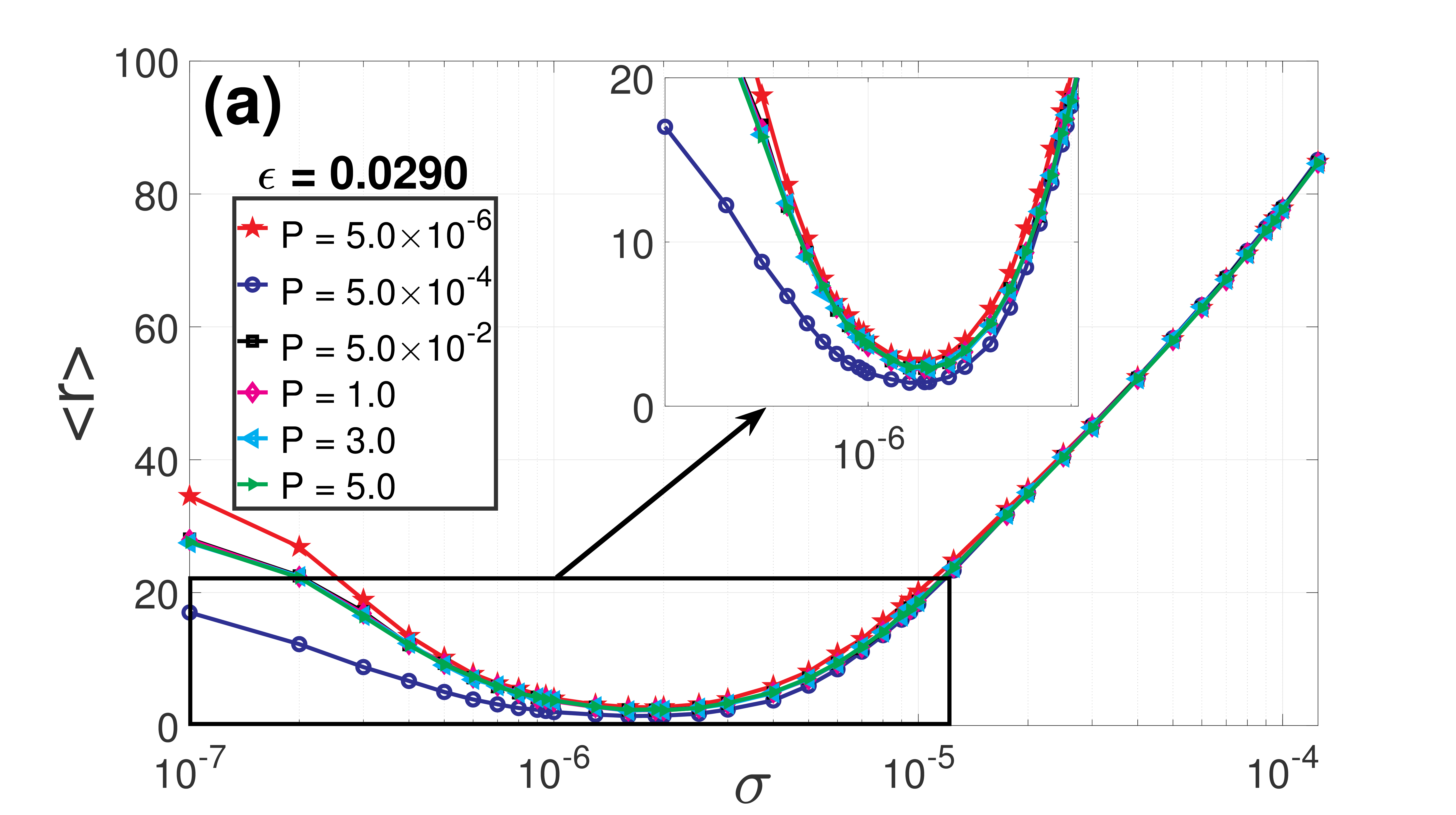}
\includegraphics[width=\columnwidth]{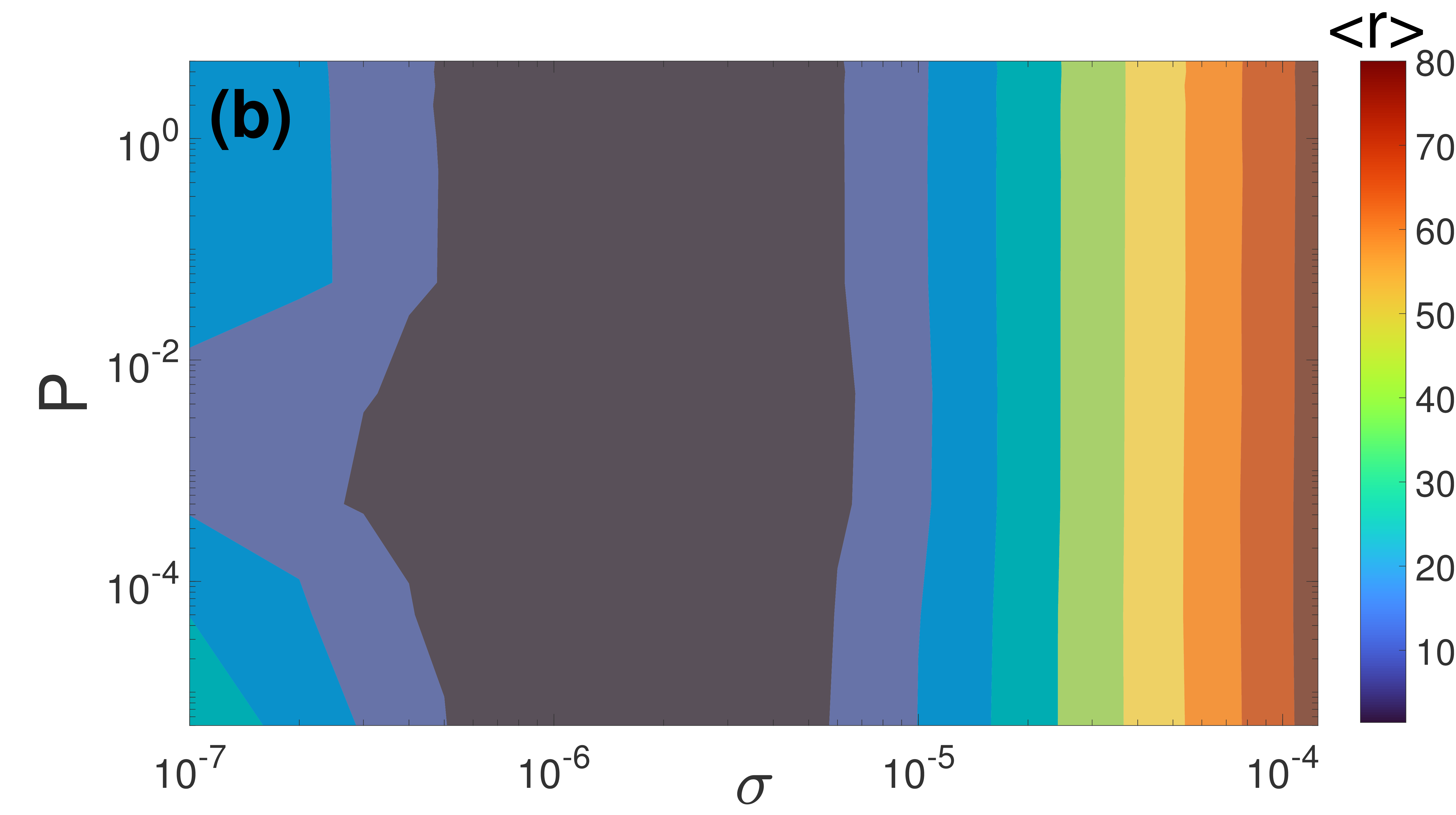}
\includegraphics[width=\columnwidth]{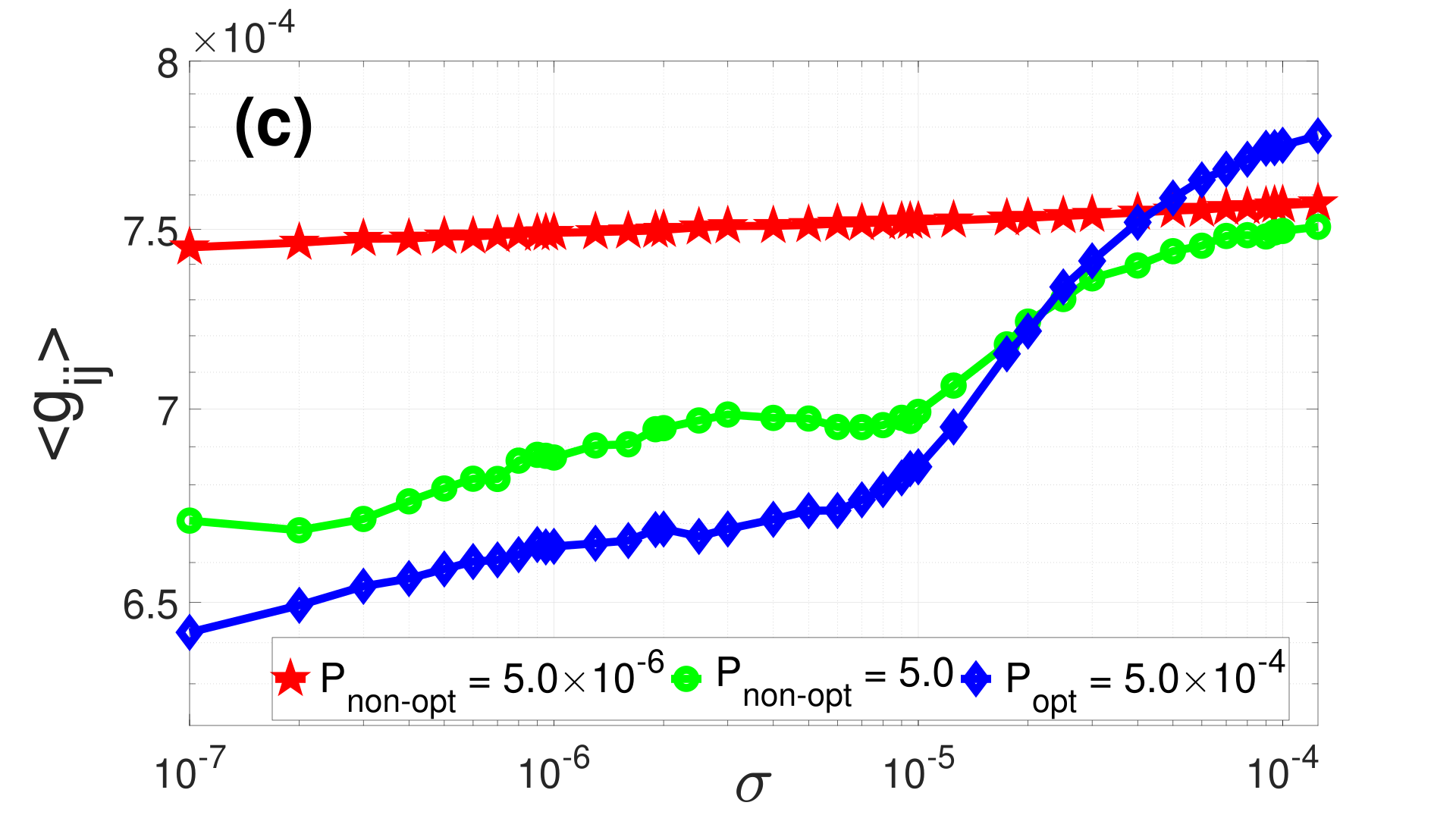}
\caption{Effect of  STDP (controlled by $P$)  on collective firing behavior of the SW network at $\varepsilon =0.0290$ in \textbf{(a)} and \textbf{(b)}. Associated population- and time-averaged synaptic weights of the network $\langle g_{ij} \rangle$ vs. noise $\sigma$ two non-optimal and the optimal values of $P$ in \textbf{(c)}.  $a=-0.05$, $b=1.0$, $c= 2.0$, $V_\text{syn}=2.0$, $V_\text{shp}=0.05$, $\beta=0.25$, $\langle k \rangle=4$, $F=0.0$, $\tau_a=\tau_b=2.0$, $B=0.5$, $A=B/P$, $N=70$.}
\label{fig:9}
\end{figure}

 We now set the rewiring frequency to $F=0$ and switch off HSP to study the effect of STDP on ISR only. Recall that STDP can be controlled by the parameter $P$, which represents the ratio between the adjusting depression $B$ and potentiation $A$ rate parameters. By fixing $B=0.5$ so that $A$ is given by $A=0.5/P$, we vary $P$ so that STDP is either depression dominated (i.e., $P>1$) or \marius{potentiation dominated} (i.e., $P<1$).  In this section of the paper, we are interested in the effects of STDP on ISR when $P$ varies in $[5.0\times10^{-6},5.0]$. Here, we also vary $\varepsilon $ and consider these effects \marius{within the interval supporting bi-stability.} The results are shown in Figs.~\ref{fig:7}-\ref{fig:9}, respectively.

In Figs.~\ref{fig:7}\textbf{(a)} and \textbf{(b)} with $\varepsilon =0.02501$, ISR is not significantly affected by changing values of $P$. This less pronounced effect of ISR as $P$ changes is an immediate consequence of the small basin of attraction of the fixed point when the timescale parameter $\varepsilon $ is near $\varepsilon _\text{HB}=0.02501$. Nevertheless, we can still see from the inset of Fig.~\ref{fig:7}\textbf{(a)} that the larger value of $P$ (see, e.g., the green curve with $P=5.0=P_\text{opt}$) induces a slightly deeper minimum of the  $\langle r \rangle$ curve when compared to smaller values of $P$, i.e., ISR is slightly enhanced.
%%%%%%%%%%%%%%%%%%%%%%%%%%%%%%
%%%%%%%%%%%%%%%%%%%%%%%%%%%%%%

In Figs.~\ref{fig:8}\textbf{(a)} and \textbf{(b)}, setting an intermediate value of $\varepsilon =0.0275375$ within the \marius{bi-meta-stable interval}, we observe that changing $P$ has a more significant effect on ISR when compared to $\varepsilon =0.02501$. Larger values of $P$ enhance ISR up to a certain threshold, upon which increasing $P$ further enhances ISR no more. More precisely, there is a significant downward shift in the ISR curve as $P$ changes from $5.0\times10^{-5}$  to $5.0\times10^{-4}$, after which ISR cannot be enhanced further.
%%%%%%%%%%%%%%%%%%%%%%%%%%%%%%
%%%%%%%%%%%%%%%%%%%%%%%%%%%%%%

In Figs.~\ref{fig:9}\textbf{(a)} and \textbf{(b)}, we fix the parameter $\varepsilon =0.0290$. Contrasting  Figs.~\ref{fig:7}\textbf{(a)} and \textbf{(b)} and Figs.~\ref{fig:8}\textbf{(a)} and \textbf{(b)}, we now observe two different behaviors: (i) increasing $P$ has a more significant effect on ISR, especially at weaker noise intensities ($\sigma\in[10^{-7},10^{-6}]$);  and (ii) the lowest ISR curve is achieved for an intermediate value of $P=5.0\times10^{-4}$ (rather than the largest value of $P=5.0$).
%%%%%%%%%%%%%%%%%%%%%%%%%%%%%%
%%%%%%%%%%%%%%%%%%%%%%%%%%%%%%

To gain further insight into the behaviors shown in Figs.~\ref{fig:7}\textbf{(a)} and \textbf{(b)}-\ref{fig:9}\textbf{(a)} and \textbf{(b)}, we computed the corresponding variations of the average synaptic weight $\langle g_{ij} \rangle$ of the network as a function of the noise intensity $\sigma$. The results are shown in Figs.~\ref{fig:7}\textbf{(c)}-\ref{fig:9} \textbf{(c)}. 
One sees that when $P$ is at its optimal value [i.e., $P_\text{opt}=5.0$ in  Figs.~\ref{fig:7}\textbf{(a)} and \ref{fig:8}\textbf{(a)}, and $P_\text{opt}=5.0\times10^{-4}$ in  Fig.~\ref{fig:9}\textbf{(a)}], the values of the average synaptic weight $\langle g_{ij} \rangle$ in the intervals of the noise intensity in which lowest ISR curves are achieved [i.e., $\sigma\in(1.25\times10^{-5}, 4.0\times10^{-5})$, $\sigma\in(3.0\times10^{-6},1.25\times10^{-5})$, and $\sigma\in(5.0\times10^{-7},6.0\times10^{-6})$ in Figs.~\ref{fig:7}\textbf{(a)}-~\ref{fig:9}\textbf{(a)}, respectively] are lower than those computed at the non-optimal values of $P$. The behavior can be explained by the fact that larger (smaller) values of average synaptic weight $\langle g_{ij} \rangle$ induced by the largest (smallest) value of $P$ in Figs.~\ref{fig:7}\textbf{(a)}-\ref{fig:8}\textbf{(a)} and an intermediate value of $P$ in Fig.\ref{fig:9}\textbf{(a)}] establishes a stronger (weaker) coupling between spiking and quiescent neurons, thereby enhancing (inhibiting) the recruitment of quiescent neurons into the spike state leading to the deterioration (improvement) of ISR.

\subsubsection{Combined effects of STDP and HSP on ISR}
Our previous investigations indicated that for intermediate values near $\varepsilon =0.0275375$ within the \marius{bi-meta-stable interval} and intermediate values of the noise intensity [i.e., $\sigma\in(3.0\times10^{-6},1.25\times10^{-5})$], the effect of the HSP parameter $F$ and the STDP parameter $P$ on ISR becomes significant compared to the lower and higher values of $\varepsilon $. In particular, it is seen in Figs.~\ref{fig:6}\textbf{(b)} and \ref{fig:8}, where $\varepsilon =0.02753755$, that increasing $F$ nd $P$, respectively,  lowers the $\langle r \rangle$  curves in the noise interval $\sigma\in(3.0\times10^{-6},1.25\times10^{-5})$, indicating an enhancement of ISR in each case.

\begin{figure}
\centering
\includegraphics[width=\columnwidth]{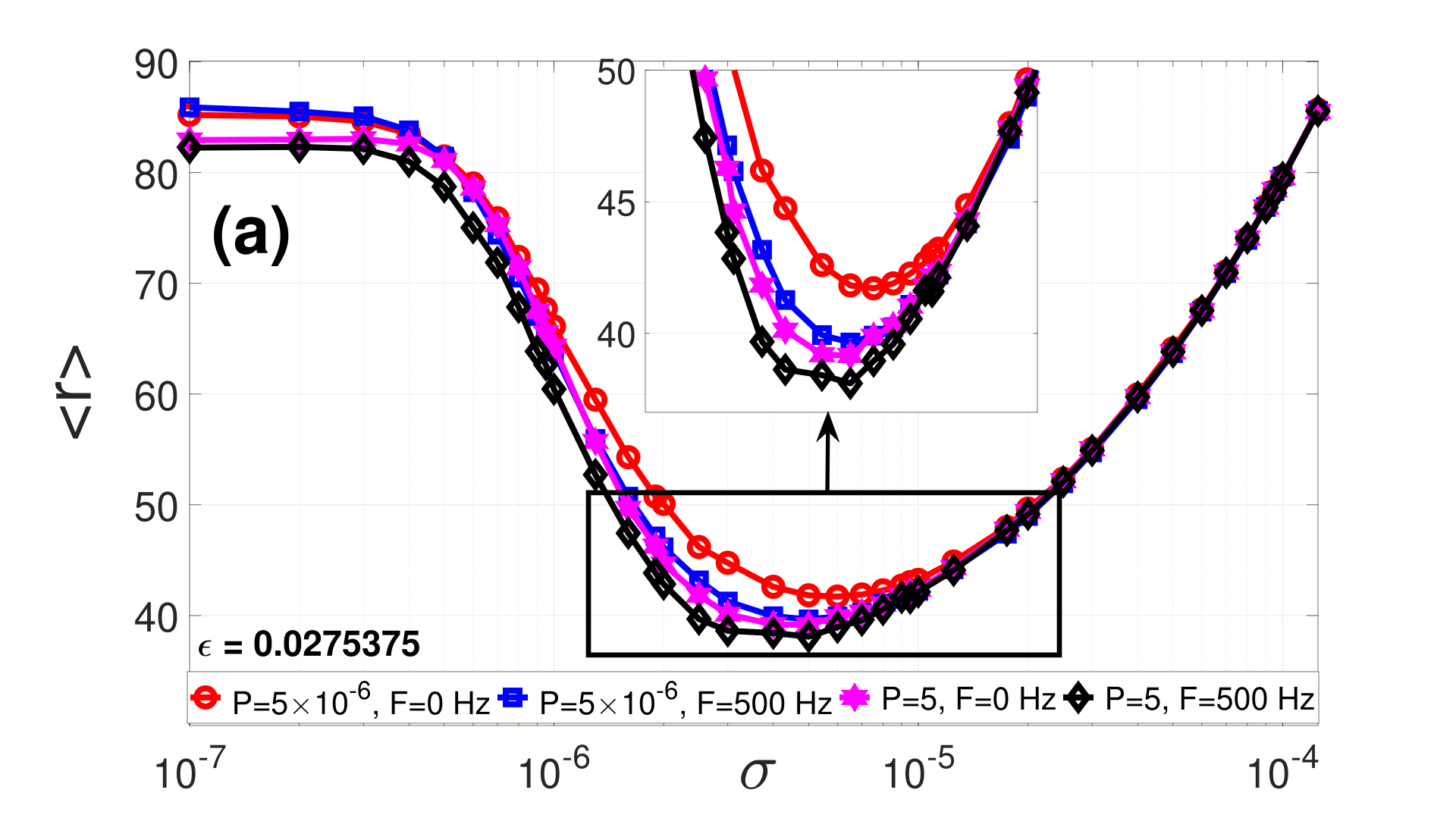}
\includegraphics[width=\columnwidth]{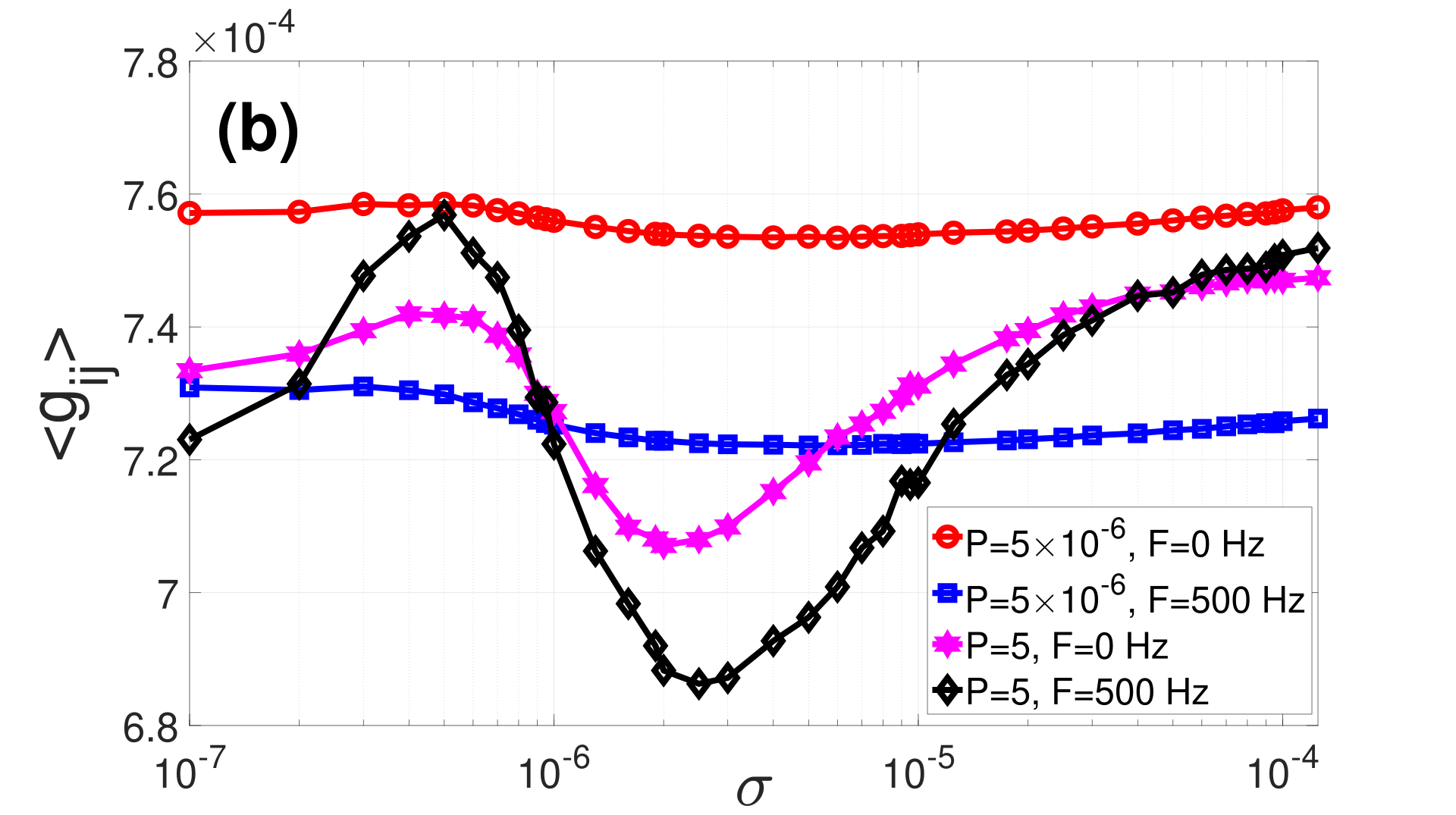}
\caption{Combined effect of  STDP and HSP on collective firing behavior of the SW network in \textbf{(a)}. Associated population- and time-averaged synaptic weights of the network $\langle g_{ij} \rangle$ vs. noise $\sigma$ for various combinations of $P$ and $F$ in \textbf{(b)}.  $a=-0.05$, $b=1.0$, $c= 2.0$, $V_\text{syn}=2.0$, $V_\text{shp}=0.05$, $\beta=0.25$, $\langle k \rangle=4$, $\tau_a=\tau_b=2.0$, $B=0.5$, $A=B/P$, $N=70$.}
\label{fig:10}
\end{figure}

A natural question arises: 
Can increasing the HSP and STDP control parameters jointly enhance ISR beyond the level achieved when only one of these parameters is increased?
And, if so, which of these parameters has the greater impact on enhancing ISR? To answer this question, in Fig.~\ref{fig:10}\textbf{(a)}, we examine the joint effect of STDP and HSP on ISR for $\varepsilon =0.0275375$. It can be seen that the deepest $\langle r \rangle$ curve is achieved when $F$ and $P$ are at their largest values --- compare the black $\langle r \rangle$ curve, with $P=5.0$ and $F=500$Hz, to the rest of the other $\langle r \rangle$  curves.  Clearly, increasing $F$ and $P$ can enhance ISR beyond the level of enhancement induced when just one of these parameters is increased to a larger value. 

Furthermore, in Fig.~\ref{fig:10}\textbf{(a)}, the separation between the minimum $\langle r \rangle_\text{min}$ of the red curve (with $P=5.0\times10^{-6}$ and $F=0$Hz) and the minimum $\langle r \rangle_\text{min}$ of the pink curve (with $P=5.0$ and $F=0$Hz) is 3.61.  While the separation between the minimum $\langle r \rangle_\text{min}$ of the red curve and the minimum $\langle r \rangle_\text{min}$ of the blue curve (with $P=5.0\times10^{-6}$ and $F=500$Hz) is 2.09. Hence, increasing the STDP control parameter $P$ has a stronger effect on ISR than increasing the HSP control parameter $F$. 

To obtain deeper insight into the behaviors depicted in Figs.\ref{fig:10}\textbf{(a)}, we calculated the corresponding changes in the average synaptic weight $\langle g_{ij} \rangle$ of the network as a function of noise intensity $\sigma$. The results are shown in Fig.~\ref{fig:10}\textbf{(b)}. We observe that the noise interval, $\sigma \in (3.0 \times 10^{-6}, 1.25 \times 10^{-5})$, where the mean firing rate $\langle r \rangle$ curves in Fig.~\ref{fig:10}\textbf{(a)} reach their minimum, corresponds to the same noise interval in which the average synaptic weight $\langle g_{ij} \rangle$ curves attain their lowest value. Specifically, within the noise range $\sigma \in (3.0 \times 10^{-6}, 1.25 \times 10^{-5})$, a decrease in the mean firing rate $\langle r \rangle$ curve corresponds to a dip in the average synaptic weight $\langle g_{ij} \rangle$ curve --- the deeper the $\langle r \rangle$ curve, the stronger the dip of the $\langle g_{ij} \rangle$ curve. This behavior can again be attributed to the fact that a weaker average synaptic weight $\langle g_{ij} \rangle$ inhibits the recruitment of quiescent neurons into the spiking state. In addition, because the basin of attraction of the stable fixed point is significantly larger (i.e., $\varepsilon _\text{HB}<\varepsilon =0.0275375$), once neurons in the spiking state are pushed into the resting state, they tend to remain at rest (at least for a very long time), thereby enhancing ISR.

\marius{At this point, it is worth mentioning that the numerical simulations presented in this paper with a small-world network of size $N$ and average degree $\langle k \rangle$, incorporating STDP and/or HSP have also been implemented (i) with different network sizes $N$ and average degrees $\langle k \rangle$, and (ii) with an STDP-driven random neural network that adheres to its randomness at all times via a different HSP rule. All the numerical simulations (not shown)  yielded qualitatively similar results to those presented in the small-world neural network.} To generate a time-varying random network topology (also generated with the Watts-Strogatz algorithm \cite{strogatz2001exploring} for $\beta=1$) while keeping the statistical network structure constant (i.e., $\beta$), we implement the following process during the rewiring of synapses: During each integration time step $dt$, if there is a synapse between neuron $i$ and $j$, it will be rewired such that neuron $i$ ($j$) connects to any other neuron except for neuron $j$ ($i$) with a probability of $\big(1-\frac{\langle k\rangle}{N-1}\big)Fdt$.

%%%%%%%%%%%%%%%%%%%%%%%%%%%%%%

\section{\label{Sec. V}Summary and conclusions}
In summary, we have conducted a numerical investigation into the phenomenon of inverse stochastic resonance (ISR) in a single FitzHugh-Nagumo (FHN) neuron and an adaptive small-world network of FHN neurons, under the influence of spike-time-dependent plasticity (STDP) and homeostatic structural plasticity (HSP). Through a combination of bifurcation analysis and numerical simulations, we identified the specific parameter values and intervals \marius{for which the individual neurons in the small-world network exhibit bi-stability, characterized by the co-existence of a stable fixed point and a limit cycle.}

The degree of ISR was shown to be highly dependent on the value of the timescale separation parameter $\varepsilon $ of the model within the \marius{bi-meta-stable interval}. Using the mean firing rate to gauge the degree of ISR, we found that as $\varepsilon $ approaches the Hopf bifurcation threshold—the lower bound of the bi-stability interval—ISR becomes less pronounced. Conversely, ISR becomes more pronounced as $\varepsilon $ moves further away from this threshold. 

Our computations demonstrated that, within the \marius{bi-meta-stable interval}, the effects of STDP and HSP on the degree of ISR vary depending on both the value of $\varepsilon $ and the interval of synaptic noise intensity $\sigma$. The effects of STDP and HSP are less significant when $\varepsilon $ is close to the lower bound of the bi-stability interval and become more significant for larger values of $\varepsilon $ within this interval and intermediate values of the synaptic noise intensity. As $\varepsilon $ gets closer to the upper bound of this bi-stability interval, ISR becomes stronger, especially at weaker synaptic noise intensities. The reason why ISR is enhanced as $\varepsilon $ is increased (within the bi-stability interval) is attributed to the fact that the basin of attraction of the stable fixed point (limit cycle) grows (shrinks) as $\varepsilon $ increases, thereby inhibiting the spiking activity of the neurons, at least for a very long time, especially at weak noise intensities.

More specifically, our results indicated that at intermediate values of the timescale parameter $\varepsilon $, increasing the rewiring frequency $F$ for HSP may noticeably enhance the degree of ISR at intermediate synaptic noise intensities. When the timescale parameter $\varepsilon $ is even closer to the upper bound of the bi-stability interval, augmenting $F$ enhances ISR, especially at weak noise intensities. Our rationale behind this behavior is that the fast rewiring of synapses constitutes a secondary source of noise (controlled by $F$), which --- in addition to the synaptic noise --- helps inhibit neural spiking and enhance ISR. Furthermore, our results indicated that at intermediate values of timescale parameter $\varepsilon $, increasing the STDP control parameter $P$ (which determines whether the neural network exhibits a potentiation- or depression-dominated in terms of the population- and time-averaged synaptic weights of the network), noticeably enhances the effect of ISR within intermediate noise intensities. When the timescale parameter $\varepsilon $ is closer to the upper bound of the bi-stability interval, the intermediate value of $P$ noticeably enhances ISR the most for weak noise intensities. Furthermore, our simulations indicated that the STDP control parameter $P$ has a greater ISR enhancement capability than the HSP control parameter $F$. 

\marius{
It is important to note that the results presented in our work may be sensitive to the choice of rewiring strategies used in the HSP rule to maintain the small-world properties of the network. Our strategy may be just one among many possible alternatives. The crucial requirement is that a homeostatic rewiring rule should ensure that the network's small-worldness is always preserved at the rewiring step. Nevertheless, our results suggest that inherent background noise, the timescale difference between the membrane potential and recovery current variables of neurons, and the prevalent STDP and HSP, can collectively play a significant role in enhancing ISR, which in turn facilitates optimal information transfer between the input and output spike trains of neurons \cite{Buchin2016PlosCB}. Interesting directions for future research are to (i) investigate and explain along the lines of Refs. \cite{Torres2020theoretical,uzuntarla2017inverse}, how STDP and HSP affect the size and depth of the basins of attraction of the meta-stable limit cycle and the meta-stable fixed point, which then affects ISR; and (ii) investigate how ISR is affected when the HSP rule considered in this paper also depends on the spike activity of the connected neurons, rather than solely on the structural configuration of the network topology.}

\marius{Adaptivity has also been considered in other forms in the literature. For example, Bačić {\it et al.}~\cite{bavcic2018inverse} investigated the adaptive coupling in excitable active rotators, wherein the plasticity plays two roles in constructing ISR, i.e., generating multistability and guiding the equilibria to be the focus, both facilitating the emergence of ISR (\Jinjie{see also \cite{bavcic2020two}}). By considering the dynamic synapses that feature short-term synaptic plasticity, Uzuntarla {\it et al.}~\cite{2017double} showed that the interval of presynaptic firing rate over which ISR exists can be extended or diminished. They further found a novel phenomenon termed double ISR under proper biophysical conditions. These researches, together with the findings obtained in this work, show that adaptivity may be beneficial to realistic biological systems within various ISR regimes.}

An experimental study has demonstrated that ISR can lead to a locally optimal information transfer between the input and output spike trains of the Purkinje neurons \cite{Buchin2016PlosCB}.
Recent experiments have shown that signaling molecules, such as acetylcholine, can modulate STDP \cite{brzosko2019neuromodulation}. Advances in experimental neuroscience have facilitated the manipulation of synaptic control in the brain through drugs that affect neurotransmitters \cite{pardridge2012drug} or by using optical fibers to stimulate genetically engineered neurons selectively \cite{packer2013targeting}. Our findings yield practical implications in locally enhancing optimal information transfer between input and output spike trains via ISR in both experimental settings and artificial neural circuits. 
Thus, our study prompts exciting venues for further experimental work on ISR in neural networks.
Finally, our findings suggest the possibility of developing ISR enhancement strategies for noisy artificial neural circuits, aimed at optimizing local information transfer between input and output spike trains in neuromorphic systems.
%%%%%%%%%%%%%%%%%%%%%%%%%%%%%%%%%%%%%%%%%%%%%%%%%%%%%%%%%%%%%%%%%%%%%%%%%%%%%%%%%%%%%%%%%%%%%%%%%%%%%%%%%%%

\begin{acknowledgments}
M.E.Y. acknowledges the support from the Deutsche Forschungsgemeinschaft (DFG, German Research Foundation) -- Project No. 456989199. J.Z. acknowledges the support from the National Natural Science Foundation of China (Grant No. 12202195). M.E.Y. and E.A.M. gratefully acknowledge financial support from the Royal Swedish Physiographic Society of Lund, Sweden.
\end{acknowledgments}

\section*{Author Declarations}
\subsection*{Conflict of Interest}
The authors have no conflicts to disclose.

\section*{Data Availability Statement}  
The simulation data supporting this study's findings are available from the corresponding author upon reasonable request.

% \appendix
% \section{Appendixes}

\begin{widetext}
%%%%%%%%%%%%%%%%%%%%%%%%%%%%%%%%%%%%%%%%%%%%%%%%%%%%%%%%%%%%%%%%%%%%%%%%%%%%%%%%%%%%%%%%%%%%%%%%%%%%%%%%%%%%%%%%%%%%%%%%%%%%%%%%%%%%%%%%%%%%%%%%%%%%%
\section{\label{Sec. VI} appendix}
\begin{algorithm}[H]
\tcc{ 
$X_i(t)$ $=\{ v_i(t),\:w_i(t)\}$: variables of coupled SDEs in Eq.\eqref{eq:1} \;
$t$: time\;
$T$: total integration time\;
$T_0$: Transient time\;
$N$: network size\;
$B$: adjusting depression rate parameter\;
$\beta$: rewiring probability in Watts-Strogatz algorithm\;
$R$: number of realizations\;
$F$: HSP control parameter (rewiring frequency of synapses)\;
$P$: STDP control parameter\;
$P_\text{min}$: min of $P$\;
$P_\text{max}$: max of $P$\;
$\varepsilon $: timescale separation parameter\;
$\sigma$: synaptic noise intensity\;
$\sigma_\text{min}$: min of $\sigma$\;
$\sigma_\text{max}$: max of $\sigma$\;
$\ell_{ij}(t)$: adjacency  matrix\;
$n_\text{spike},i,m$ : number spikes of the $i^{th}$ neuron at the $m^{th}$ realization\;
$\langle  r \rangle_{i,m}$: meaning firing rate of the $i^{th}$ neuron at the $m^{th}$ realization\;
$\langle  r \rangle$: average of the meaning firing rate over $R$\;
$g_{ij}(t)$: synaptic weights\; 
$\langle  g_{ij} \rangle_\text{m}$: average of synaptic weights over $t$ and $N$ of the $m^{th}$ realization\;
$\langle g_{ij} \rangle$: average of mean synaptic weights over $R$\;
}	
\KwInput{$T$, $T_0$, $N$, $R$, $F$, $P$, $\beta$}
\KwOutput{$\langle r \rangle, \langle g_{ij} \rangle$}
$P \gets P_\text{min}$ \tcp*{Initialize the adjusting rate parameter}
\While{$ P \leq P_\text{max}$ }  { 
$\sigma \leftarrow \sigma_\text{min}$ \tcp*{Initialize the timescale parameter}
	\While{$ \sigma \leq \sigma_\text{max}$ } {
		\For{$m \in  1,2,\dots,R$}{
			Init $X_i(t)\:,\:\ell_{ij}(t)$ \tcp*{Random initial conditions of SDEs and initial SW network adjacency matrix}
			\For{$t \in 0,\dots,T $}{
				Integrate network of SDEs in Eqs.~\eqref{eq:1} \tcp*{Using the Euler-Maruyama method}
				Record the current voltage spike times $t^{n}_i$ from $V_i(t)$\tcp*{Times $t$ at which $v_i(t)\ge v_{\mathrm{th}}=0.25$}
				\If{$\Delta t_{ij}:=t_i - t_j > 0$}  {
					$\Delta M\gets \displaystyle{\frac{B}{P}}\exp{(-\lvert\Delta t_{ij}\rvert/\tau_{p})}$ 
			\tcp*{$t_i$\:,\:$t_j$: nearest-spike times of post ($i$) \& pre ($j$) neuron} 
			}
		
				\If{$\Delta t_{ij} <0$} {
				$\Delta M \gets  -B\exp{(-\lvert\Delta t_{ij}\rvert/\tau_{d})}$ 
			}
		
			\If{$\Delta t_{ij} = 0$} {
				$\Delta M \gets  0$ 
			}
   
				$g_{ij}(t) \gets g_{ij}(t) + g_{ij}(t)\Delta M$\tcp*{update synaptic weights}
					$\ell_{ij}(t) \gets \widetilde{\ell_{ij}}(t) $ \tcp*{Update the adjacency matrix with  $\widetilde{\ell_{ij}}(t)$ obtained by randomly rewiring $\ell_{ij}(t)$ with frequency $F$ according rewiring algorithm that preserve the small-worldness of the SW network generated with rewiring probability $\beta$.}

      \If{$t\geq T_0$}{
		$r_{i,m} \gets \displaystyle{n_{\text{spike},i,m}}$ \:,
			$\langle g_{ij} \rangle_\text{m}  \gets \displaystyle{\bigg \langle N^{-2}\sum\limits_{i=1}^{N}\sum\limits_{j=1}^{N}g_{ij}(t)\bigg \rangle_t}$\;
             }
		}
			Add $r_{i,m}$ to $r_i$\;
			Add $\langle g_{ij} \rangle_\text{m}$ to $\overline{\langle g_{ij} \rangle}$\;
		}
            $\langle  r_i \rangle \gets r_i[R(T-T_0)]^{-1}$ \tcp*{Compute averages over the $R$ realizations}
             Add $\langle  r_i \rangle$  to $\overline{\langle  r \rangle}$\;
            $\langle  r \rangle \gets \overline{\langle  r \rangle}/N$   \tcp*{Compute averages over the $R$ realizations}
		  $\langle g_{ij} \rangle \gets \overline{\langle g_{ij} \rangle}/R$ \tcp*{Compute averages over the $R$ realizations}
		 $ \sigma \gets \sigma + \Delta \sigma$ \tcp*{Increment the noise intensity}
	} 
$P \gets P + \Delta P$  \tcp*{Increment the STDP control parameter}
\caption{Flow of control in the simulations of Fig.\ref{fig:7}. Adapted for the simulations in the rest of the figures.}
}
\end{algorithm}
%%%%%%%%%%%%%%%%%%%%%%%%%%%%%%%%%%%%%%%%%%%%%%%%%%%%%%%%%%%%%%%%%%%%%%%%%%%%%%%%%%%%%%%%%%%%%%%%%%%%%%%%%%%%%%%%%%%%%%%%%%%%%%%%%%%%%%%%%%%%%%%%%%%%%
\end{widetext}
%%%%%%%%%%%%%%%%%%%%%%%%%%%%%%%%%%%%%%%%%%%%%%%%%%%%%%%%%%%%%%%%%%%%%%%%%%%%%%%%%%%%%%%%%%%%%%%%%%%%%%%%%%%%%%%%%%%%%%%%%%%%%%%%%%%%%%%%%%%%%%%%%%%%%

%
%\bibliography{ref.bib}% Produces the bibliography via BibTeX.

%merlin.mbs aipnum4-1.bst 2010-07-25 4.21a (PWD, AO, DPC) hacked
%Control: key (0)
%Control: author (8) initials jnrlst
%Control: editor formatted (1) identically to author
%Control: production of article title (0) allowed
%Control: page (1) range
%Control: year (1) truncated
%Control: production of eprint (0) enabled
\providecommand{\noopsort}[1]{}\providecommand{\singleletter}[1]{#1}%\providecommand{\noopsort}[1]{}\providecommand{\singleletter}[1]{#1}%

\end{document}